\documentclass[a4paper,11pt]{article}
\pdfoutput=1 
\usepackage{jheppub} 
\usepackage{amssymb}
\usepackage{epsfig}
\usepackage{graphicx}%
\usepackage{amsmath}%

\title{\boldmath Quantitative Study of Geometrical Scaling \\
in Deep Inelastic Scattering at HERA}
\author{Michal Praszalowicz}
\author{and Tomasz Stebel}
\affiliation{M. Smoluchowski Institute of Physics, Jagiellonian University,\\
Reymonta 4, 30-059 Krak{\'o}w, Poland}

\emailAdd{michal@if.uj.edu.pl}
\emailAdd{tomasz.stebel@uj.edu.pl}

\abstract{We propose a method to assess the quality of geometrical scaling in Deep
Inelastic Scattering and apply it to the combined HERA data on $\gamma^{\ast
}p$ cross-section. Using two different approaches based on Bjorken $x$ binning
and binning in $\gamma^{\ast}p$ scattering energy $W$, we show that
geometrical scaling in variable $\tau\sim Q^{2} x^{\lambda}$ works well up to
Bjorken $x$'s 0.1. The corresponding value of exponent
$\lambda$ is 0.32 -- 0.34.}

\begin{document}
\maketitle
\flushbottom

\section{Introduction}

\label{intro}

Geometrical scaling (GS) introduced in Ref.~\cite{Stasto:2000er} in the
context of low $x$ Deep Inelastic Scattering (DIS) is by now well established
phenomenon attributed to the existence of an intermediate energy scale, called
saturation scale $Q_{\text{s}}(x)$. Saturation scale is defined as the border
line between dense and dilute gluonic systems within a proton (for review
see \emph{e.g.} Refs.~\cite{Mueller:2001fv,McLerran:2010ub}). GS has been also
observed in DIS on nucleus \cite{eA} and in diffraction \cite{Marquet:2006jb}.
Recently it has been shown that GS is also exhibited by the $p_{\text{T}}$
spectra at the LHC \cite{McLerran:2010ex,Praszalowicz:2011tc,Praszconf}.
Geometrical scaling has a natural explanation within the theory of saturation
and the Color Glass Condensate
\cite{Gribov:1984tu,Mueller:1985wy,McLerran:1993ni,Ayala:1995hx}.
However, one should note that GS extends well above the saturation scale
both in the DGLAP \cite{Kwiecinski:2002ep} and BFKL \cite{Iancu:2002tr} 
evolution schemes. It has been argued in Ref.~\cite{Caola:2008xr} that the appearance 
of GS in course of the evolution is fairly independent of the initial conditions.

Geometrical scaling takes place if some observable that in principle depends
on two independent kinematical variables like $Q^{2}$(or $p_{\text{T}}^{2}$)
and $W$ (\emph{i.e.} $\gamma^{\ast}p$ c.m.s. energy) depends only on a
specific combination of them, namely on%
\begin{equation}
\tau=\frac{Q^{2}}{Q_{\text{s}}^{2}(x)} \label{tau}%
\end{equation}
called scaling variable. Here
\begin{equation}
x=\frac{Q^{2}}{Q^{2}+W^{2}-M_{\text{p}}^{2}} \label{xdef}%
\end{equation}
is Bjorken $x$ variable, $M_p$ stands for the proton mass,
and the saturation scale $Q_{\text{s}}$ takes the
following form \cite{Stasto:2000er,GolecBiernat:1998js}%
\begin{equation}
Q_{\text{s}}^{2}(x)=Q_{0}^{2}\left(  \frac{x}{x_{0}}\right)  ^{-\lambda}.
\label{Qsat}%
\end{equation}
Here $Q_{0}$ and $x_{0}$ are free parameters which can be fitted to the data
within some specific model of DIS, and exponent $\lambda$ is a dynamical
quantity of the order of $\lambda\sim0.3$. Physical observable that exhibits
GS in the case of DIS is $\gamma^{\ast}p$ cross-section $\sigma_{\gamma^{\ast
}p}(x,Q^{2})=4\pi^{2} \alpha_{\mathrm{em}} F_{2}(x,Q^{2})/Q^{2}$.

Although many authors, following the original paper by Stasto, Golec-Biernat
and Kwecinski \cite{Stasto:2000er}, have shown that GS is seen in DIS data,
until now there was no quantitative, model independent analysis of its
applicability domain. Different forms of scaling variable have been tested in
a series of papers \cite{Gelis:2006bs,Beuf:2008bb,Royon:2010tz} where the so
called Quality Factor (QF) has been defined and used as a tool to assess the
quality of geometrical scaling. These authors, however, constrained their
analysis only to the domain of small Bjorken $x$'s, $x<0.01$ and concentrated
on testing the different forms of scaling variable $\tau$. The QF is a new
tool for which the quantitative statements, like the one concerning the value
of exponent $\lambda$ for example, are not based on the standard chi-square analysis.

In contrast, in this paper we propose a new numerical criterion for GS which
serves as a tool to extract exponent $\lambda$ by standard $\chi^{2}$
minimization. We consider $\sigma_{\gamma^{\ast}p}(x_{i},Q^{2})$ for different
fixed $x_{i}$'s as functions of $Q^{2}$. Geometrical scaling hypothesis means
that%
\begin{equation}
\sigma_{\gamma^{\ast}p}(x_{i},Q^{2})=\frac{1}{Q_{0}^{2}}F(\tau)
\end{equation}
where $F(\tau)$ is a universal dimensionless function of $\tau$. Therefore if
cross-sections $\sigma_{\gamma^{\ast}p}(x_{i},Q^{2})$ for different $x_{i}$
are evaluated not in terms of $Q^{2}$ but in terms of $\tau$, they should fall
on one universal curve (see Figs.~\ref{figGSinx} and \ref{figGSinW} in
Sect.~\ref{sumout}). This in turn means that if we calculate ratio of
cross-sections for different Bjorken $x_i$'s each expressed in terms of $\tau$,
we should get unity (with an accuracy of a few percent) independently of $\tau$. 
This allows to determine power
$\lambda$ by minimizing deviations of these ratios from unity. We can form as
many ratios as there are different pairs of $x_{i}$'s in overlapping regions
of $Q^{2}$ (or more precisely of $\tau$). Needless to say that the best values
of $\lambda$'s extracted from different ratios should coincide within errors.

In an ideal case one would choose the lowest possible $x$ as the reference
$x_{\text{ref}}$ to calculate the ratios of the cross-sections. Unfortunately
there is a strong correlation between Bjorken $x$'s and values of $Q^{2}$
measured by HERA \cite{HERAcombined}, and therefore there is no single value
of $x_{i}$ which covers all available values of $Q^{2}$. It turns out that the
coverage in space of $W$ (rather than $x$) and $Q^{2}$ is much broader.
Therefore in what follows we also study the quality of GS for $\sigma
_{\gamma^{\ast}p}(W_{i},Q^{2})$ in bins of $W$ although this requires
"rebinning" of the data which are provided by HERA experiments only in bins of
$(x,Q^{2})$. There is, however, an advantage of such a procedure, as it is
almost identical to the analysis applied to the $p_{\text{T}}$ spectra in $pp$
collisions at the LHC \cite{McLerran:2010ex,Praszconf}.

Since our analysis is sensitive only to the variations of scaling variable
$\tau$ with $x$ and not to the absolute value of $\tau$, we choose in the
following (unless specified otherwise) $Q_{0}=1$ GeV/$c$ and $x_{0}=1$,
\emph{i.e.}:%
\begin{equation}
\tau=Q^{2}x^{\lambda}. \label{taudef}%
\end{equation}
The absolute value of the saturation scale can be inferred only from some
explicit model of DIS at low $x$. For the purpose of the subsequent analysis
we define $\gamma^{\ast}p$ cross-section as%
\begin{equation}
\sigma_{\gamma^{\ast}p}(x,Q^{2})=\frac{F_{2}(x,Q^{2})}{Q^{2}}%
\end{equation}
since all proportionality constants cancel out in the ratios of the
cross-sections that are the main tool used to look for geometrical scaling in
this paper. The corresponding experimental error is therefore given by%
\begin{equation}
\Delta\sigma_{\gamma^{\ast}p}= \frac{\Delta F_{2}}{Q^{2}}
\label{error1}
\end{equation}
where experimental error of $Q^2$ is essentially negligible\footnote{We thank 
Halina Abramowicz and Iris Abt from ZEUS collaboration at HERA for 
clarification of this point.}.  

Our findings can be shortly summarized as follows. Geometrical scaling in
variable (\ref{taudef}) works well up to Bjorken $x$'s 0.1.
The corresponding value of exponent $\lambda$ is 0.32 -- 0.34.

The paper is essentially divided into two parts. In Sect.~\ref{Bjxbinning} we
discuss GS in terms of Bjorken variable $x$, whereas in Sect.~\ref{Wbinning}
in terms of $\gamma^{\ast}p$ scattering energy $W$, in both cases for $e^{+}p$
scattering. In Sect.~\ref{sumout} we summarize our results and discuss other
experiments and also HERA data for $e^{-}p$. We also present possible
applications of the method proposed in this paper to test other possible forms
of scaling variable $\tau$ and to investigate scaling of the charm cross-section.
Preliminary results and technical details can be found in Ref.  \cite{Stebel:2012ky}.

\section{Bjorken $x$ binning}

\label{Bjxbinning}

\begin{figure}[t]
\centering
\hspace{-0.7cm} \includegraphics[width=10cm]{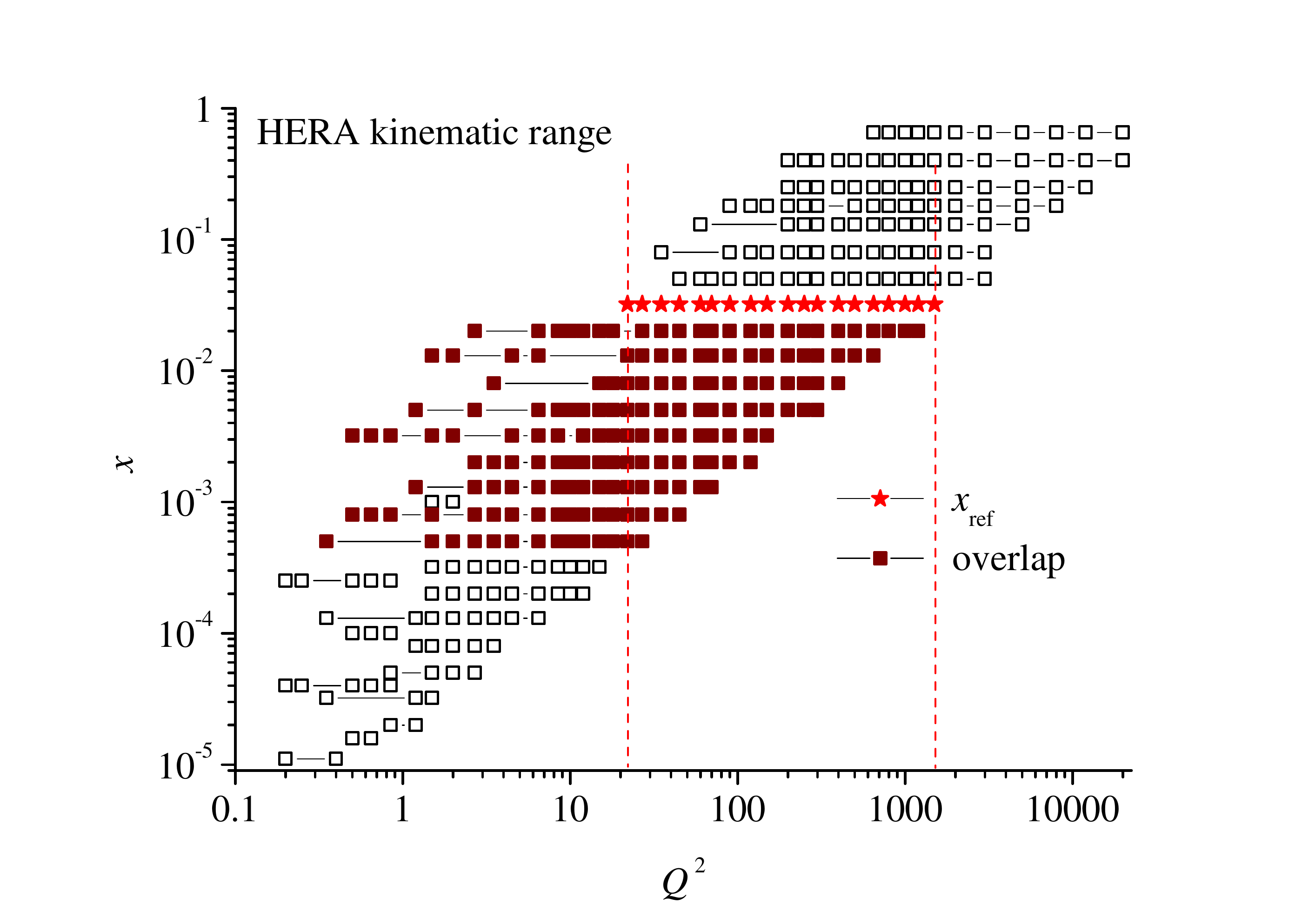}\caption{HERA
kinematic range in ($Q^{2},x$) plane. It is shown which Bjorken $x$'s (brown
 squares) are chosen for given $x_{\mathrm{ref}}$ (red
stars) to construct ratios of $\gamma^{\ast}p$ cross-sections in scaling
variable $\tau$. }%
\label{HERAxref}%
\end{figure}

Combined study of ZEUS and H1 \cite{HERAcombined} provides 432 $e^{+}p$ data
points for DIS structure function $F_{2}$ in terms of Bjorken $x$ and $Q^{2}$.
Here we shall use 348 points with Bjorken $x$'s that have at least 2 points in
$Q^{2}$ \cite{Stebel:2012ky}. This choice corresponds to $5.52\times10^{-6}\leq x\leq0.65$. Since
there is a strong correlation between $x$ and $Q^{2}$ as far as HERA kinematic
range is concerned (see Fig.~\ref{HERAxref}), the analysis outlined in
Sect.\ref{intro} requires some refinement. We apply here the following
procedure. First we choose some $x_{\mathrm{ref}}$ and consider all Bjorken
$x_{i}$'s smaller than $x_{\mathrm{ref}}$ that have at least two overlapping
points in $Q^{2}$ (or more precisely in scaling variable $\tau$), as depicted
in Fig.~\ref{HERAxref} (note, however, that when we make the same plot in
terms of variable $\tau$ rather than $Q^{2}$ the overlaps change; the
number of overlapping points gets smaller). We require that $x_{i}%
<x_{\text{ref}}$ because GS holds for small values of Bjorken $x$ and
therefore by increasing $x_{\text{ref}}$ we will be able to see violations of
GS. For $x_{i}<x_{\text{ref}}$ the ratios
\begin{equation}
R_{x_{i},x_{\text{ref}}}(\lambda;\tau_{k})=\frac{\sigma_{\gamma^{\ast}p}%
(x_{i},\tau(x_{i},Q_{k}^{2};\lambda))}
{\sigma_{\gamma^{\ast}p}(x_{\text{ref}},\tau(x_{\text{ref}%
},Q_{k,\text{ref}}^{2};\lambda))}\;\text{with}\;\tau_{k}=\tau(x_{i},Q_{k}%
^{2};\lambda)=\tau(x_{\text{ref}},Q_{k,\text{ref}}^{2};\lambda) \label{Rxdef}%
\end{equation}
are greater than 1 for $\lambda=0$. By increasing $\lambda$ one shifts
$R_{x_{i},x_{\text{ref}}}(\lambda;\tau_{k})$ towards unity with an accuracy
$\delta$ for all values of $\tau_{k}$:
\begin{equation}
R_{x_{i},x_{\text{ref}}}(\lambda;\tau_{k}) \rightarrow 1 \pm \delta.
\end{equation}
Here $\delta$ stands for theoretical accuracy of GS hypothesis for which we take 3 \%. 
GS is an approximate scaling law and with accurate combined HERA data we see its tiny violations 
which show up as an increase of ratios R with scaling variable $\tau$. Including small theoretical 
error makes our analysis immune to this effect.

Note that $\tau(x_{i},Q_{k}^{2};\lambda=0)=Q_{k}^{2}$ and ratios (\ref{Rxdef})
are essentially ratios of $\gamma^{\ast}p$ cross-sections at overlapping
values of $Q_{k}^{2}$ (as in Fig.~\ref{HERAxref}). For $\lambda\neq0$ points
of the same $Q^{2}$ but different $x$'s correspond generally to different
$\tau$'s. In order to calculate ratios (\ref{Rxdef}) at points $\tau_{k}$
corresponding to one fixed $x_{i}$, one has to interpolate the reference
cross-section 
$\sigma_{\gamma^{\ast}p}(x_{\text{ref}},\tau(x_{\text{ref}},Q^{2};\lambda))$ to
$Q_{k,\text{ref}}^{2}$ such that $\tau(x_{\text{ref}},Q_{k,\text{ref}}%
^{2};\lambda)=\tau_{k}$. Since $\gamma^{\ast}p$ cross-sections to a good
accuracy lie on straight lines as functions of $\log Q^{2}$, we apply in the
following the linear interpolation in $\log Q^{2}$. Interpolation errors 
are included in an overall error of $R_{x_{i},x_{\text{ref}}}$ (see Eq.~\ref{err1}
below).

It is clear than not all available Bjorken $x$'s can be used as
$x_{\mathrm{ref}}$. If one chooses $x_{\mathrm{ref}}$ too small then there are
no $x<x_{\mathrm{ref}}$ which have at least two points in $\tau$ (or $Q^{2}$)
within $\sigma_{\gamma^{\ast}p}(\tau(x_{\text{ref}},Q^{2}))$ domain. It turns
out that we can use only $x_{\mathrm{ref}}\geq3.2\cdot10^{-5}$.

In order to find optimal exponent $\lambda$ that minimizes deviations of
ratios (\ref{Rxdef}) from unity we form the chi-square measure%
\begin{equation}
\chi_{x_{i},x_{\text{ref}}}^{2}(\lambda)=\frac{1}{N_{x_{i},x_{\text{ref}}}-1}%
{\displaystyle\sum\limits_{k\in x_{i}}}
\frac{\left(  R_{x_{i},x_{\text{ref}}}(\lambda;\tau_{k})-1\right)  ^{2}%
}{\Delta R_{x_{i},x_{\text{ref}}}(\lambda;\tau_{k})^{2}} \label{chix1}%
\end{equation}
where the sum over $k$ extends over all points of given $x_{i}$ that have
overlap with $x_{\text{ref}}$. As already explained above, for $\lambda=0$
this is essentially the sum over all overlapping values of $Q_{k}^{2}$, for
$\lambda\neq0$ we choose measured points of $\sigma_{\gamma^{\ast}p}%
(\tau(x_{i},Q_{k}^{2};\lambda))$ and interpolate the reference cross-section
in $Q^{2}$ to the point $Q_{k,\text{ref}}^{2}$ such that $\tau(x_{\text{ref}%
},Q_{k,\text{ref}}^{2};\lambda)=\tau_{k}$.

Finally, the errors entering formula (\ref{chix1}) are calculated using%
\begin{align}
& \Delta R_{x_{i},x_{\text{ref}}}(\lambda;\tau_{k})^{2} =   \label{err1} \\
& \left( \left( 
\frac{\Delta\sigma_{\gamma^{\ast}p}(x_{i},\tau(x_{i},Q_{k}^{2}))}{
\sigma_{\gamma^{\ast}p}(x_{i},\tau(x_{i},Q_{k}^{2}))}\right)^{2} 
+\left(  \frac{\Delta\sigma_{\gamma^{\ast}p}(x_{\text{ref}},\tau(x_{\text{ref}},Q_{k,\text{ref}}%
^{2}))}%
{\sigma_{\gamma^{\ast}p}(x_{\text{ref}},\tau(x_{\text{ref}},Q_{k,\text{ref}}^2))}%
\right)  ^{2}\right) R_{x_{i},x_{\text{ref}}}(\lambda; \tau_k)^2 
+ \delta^2 & \nonumber %
\end{align}
where $\Delta\sigma_{\gamma^{\ast}p}(\tau(x,Q^{2}))$ are experimental errors
(or interpolated experimental errors with interpolation error included) 
of $\gamma^{\ast}p$ cross-sections  (\ref{error1}).

\begin{figure}[ptb]
\includegraphics[width=7.8cm,angle=0]{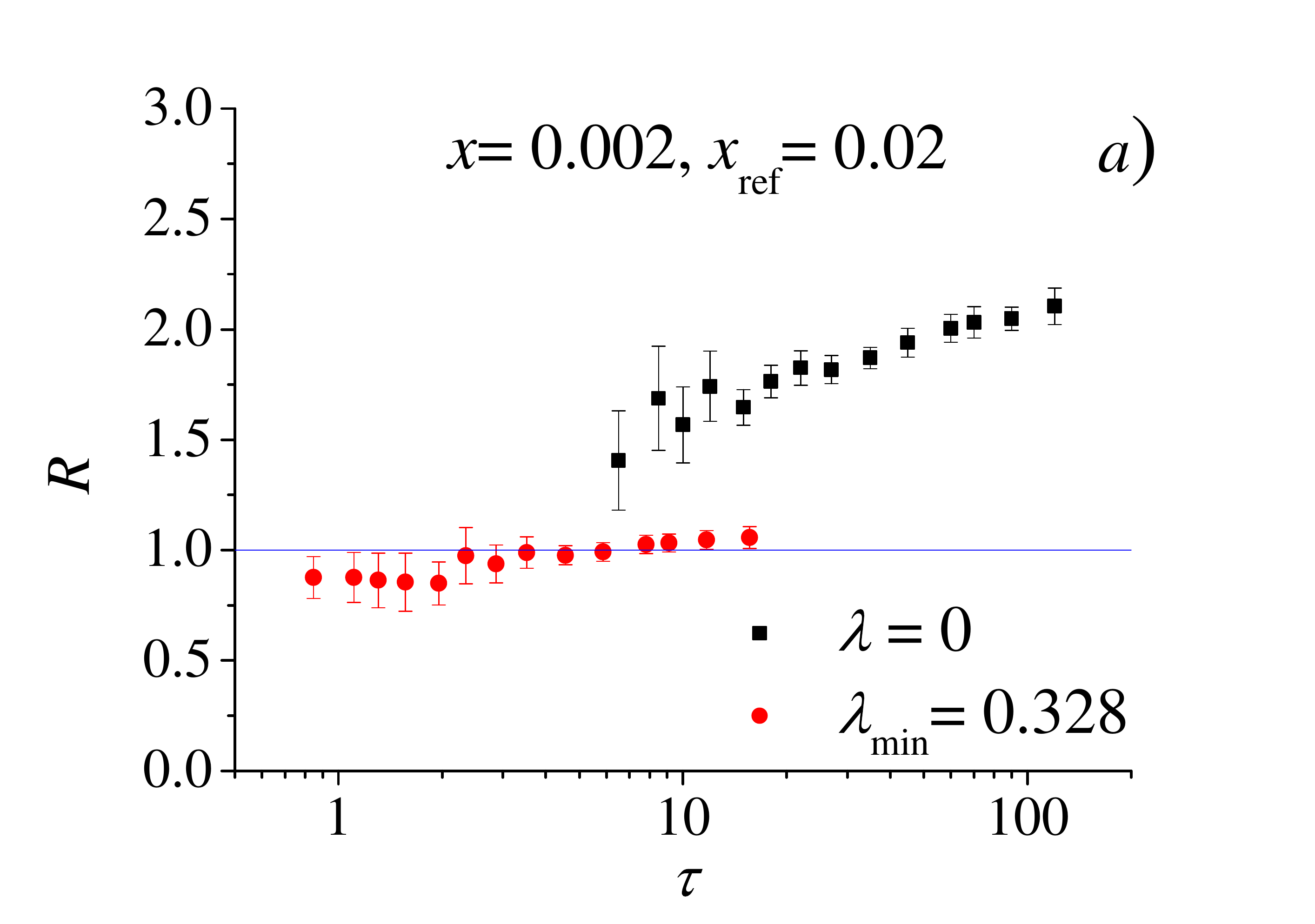}
\includegraphics[width=7.8cm,angle=0]{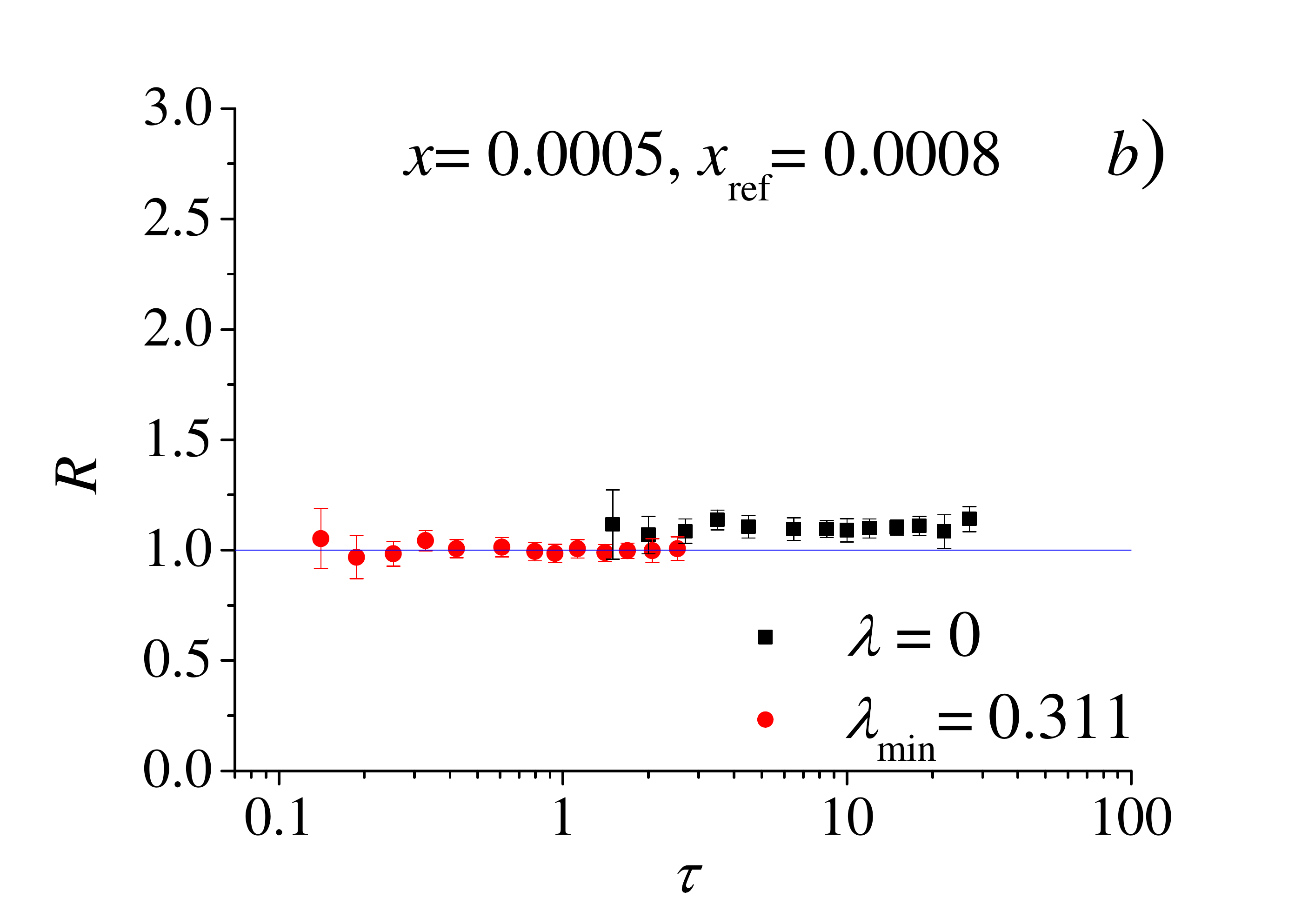}\newline%
\includegraphics[width=7.8cm,angle=0]{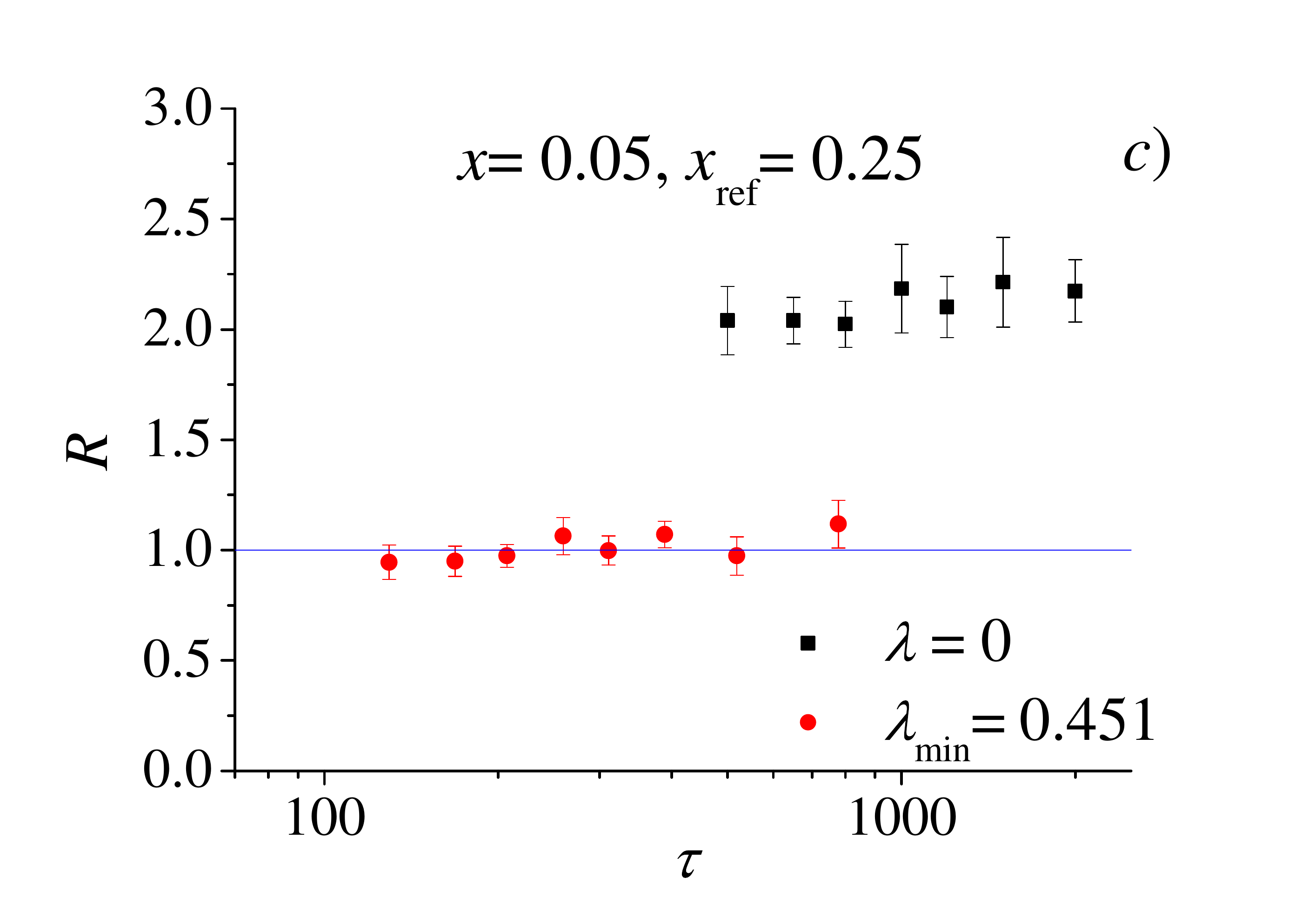}
\includegraphics[width=7.8cm,angle=0]{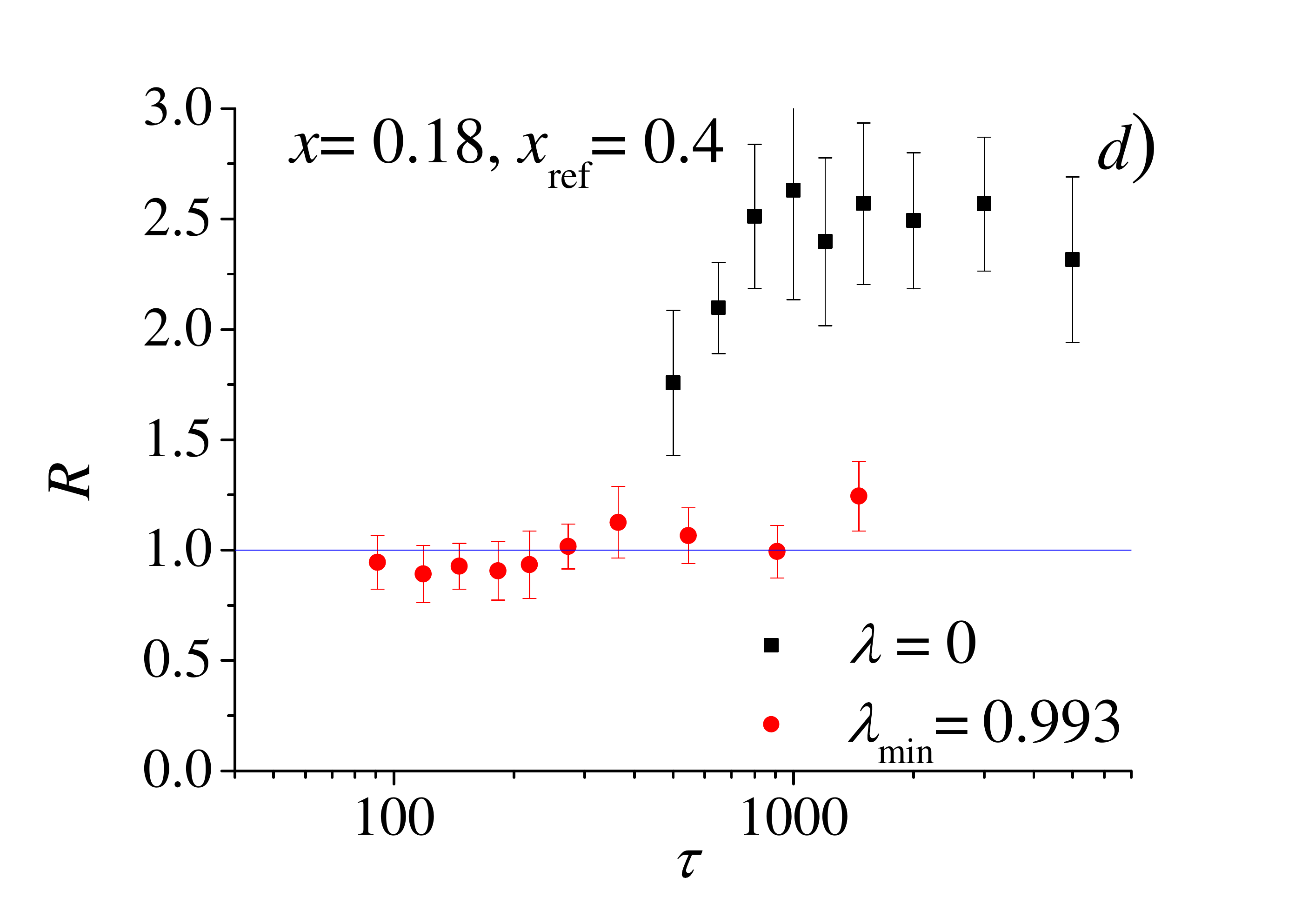}\caption{Ratios
$R_{i}(\lambda)$ for $\lambda=0$ and $\lambda=\lambda_{\mathrm{min}}$ for some
typical choices of $x$ and $x_{\mathrm{ref}}$.}%
\label{Rxxref}%
\end{figure}

By minimizing $\chi_{x,x_{\text{ref}}}^{2}(\lambda)$ of Eq.~(\ref{chix1}) with
respect to $\lambda$ one can compute the "best" value $\lambda_{\text{min}%
}(x,x_{\text{ref}})$ and its error requiring that 
$(N_{x_{i},x_{\text{ref}}}-1)\chi_{x_{i},x_{\text{ref}}}^{2}(\lambda_{\rm min})$
changes by 1 around the minimum.
In the region where GS is satisfied $\lambda_{\text{min
}}$ should not depend neither on $x$ nor on $x_{\text{ref}}$ and
$\chi_{x,x_{\text{ref}}}^{2}(\lambda_{\text{min}})$ should be small. In Fig.
\ref{Rxxref} we plot ratios $R_{x,x_{\text{ref}}}(\lambda;\tau_{k}) $ as
functions of $\tau_{k}$ for $\lambda=0$ and $\lambda=\lambda_{\text{min}}$ for
some typical values of $x$ and $x_{\text{ref}}$. In Fig.~\ref{Rxxref}.a we
plot $R$ for both $x$ and $x_{\text{ref}} $ small and quite distant,
\emph{i.e.} $x_{\text{ref}}/x=10$. Here we expect GS to be satisfied. We see
that for $\lambda=0$ ratio $R$ grows with $\tau_{k}=Q_{k}^{2}$ and is of the
order $1.5\div2$. By minimizing $\chi_{x,x_{\text{ref}}}^{2}(\lambda)$ we
obtain $\lambda_{\text{min}}=0.328$, which agrees with the expectations. In
Fig.~\ref{Rxxref}.b, both $x$ and $x_{\text{ref}}$ are small, however they are
not very distant: $x_{\text{ref}}/x=1.6$. We see that already for $\lambda=0$
ratio $R_{x,x_{\text{ref}}}(0;\tau_{k})$ is close to $1$; we can improve the
value of $\chi^{2}$ by increasing the value of $\lambda$ to $\lambda
_{\mathrm{min}}=0.311$, in fair agreement with the previous case, however one
should note that now $\chi^{2}(\lambda)$ is rather flat. Another example is
shown in Fig.~\ref{Rxxref}.c where $x_{\text{ref}}=0.25$ is relatively large
and $x$ still quite small, such that $x_{\text{ref}}/x=5$. Here again it is
possible to make $R_{x,x_{\text{ref}}}(\lambda;\tau_{k})$ close to unity,
however at the expense of rather large value of $\lambda_{\text{min}}=0.451$.
This is a clear sign of violation of the universality of exponent $\lambda$,
which we consider to be a signature of violation of geometrical scaling. Similarly
in Fig.~\ref{Rxxref}.d we show $R_{x,x_{\text{ref}}}(\lambda;\tau_{k}) $ for
both $x$ and $x_{\text{ref}}$ large and not so close: $x_{\text{ref}}/x=2.22$.
Here $\lambda_{\text{min}}=0.993$, which is outrageously large.

\begin{figure}[ptb]
\centering
\includegraphics[width=10cm,angle=0]{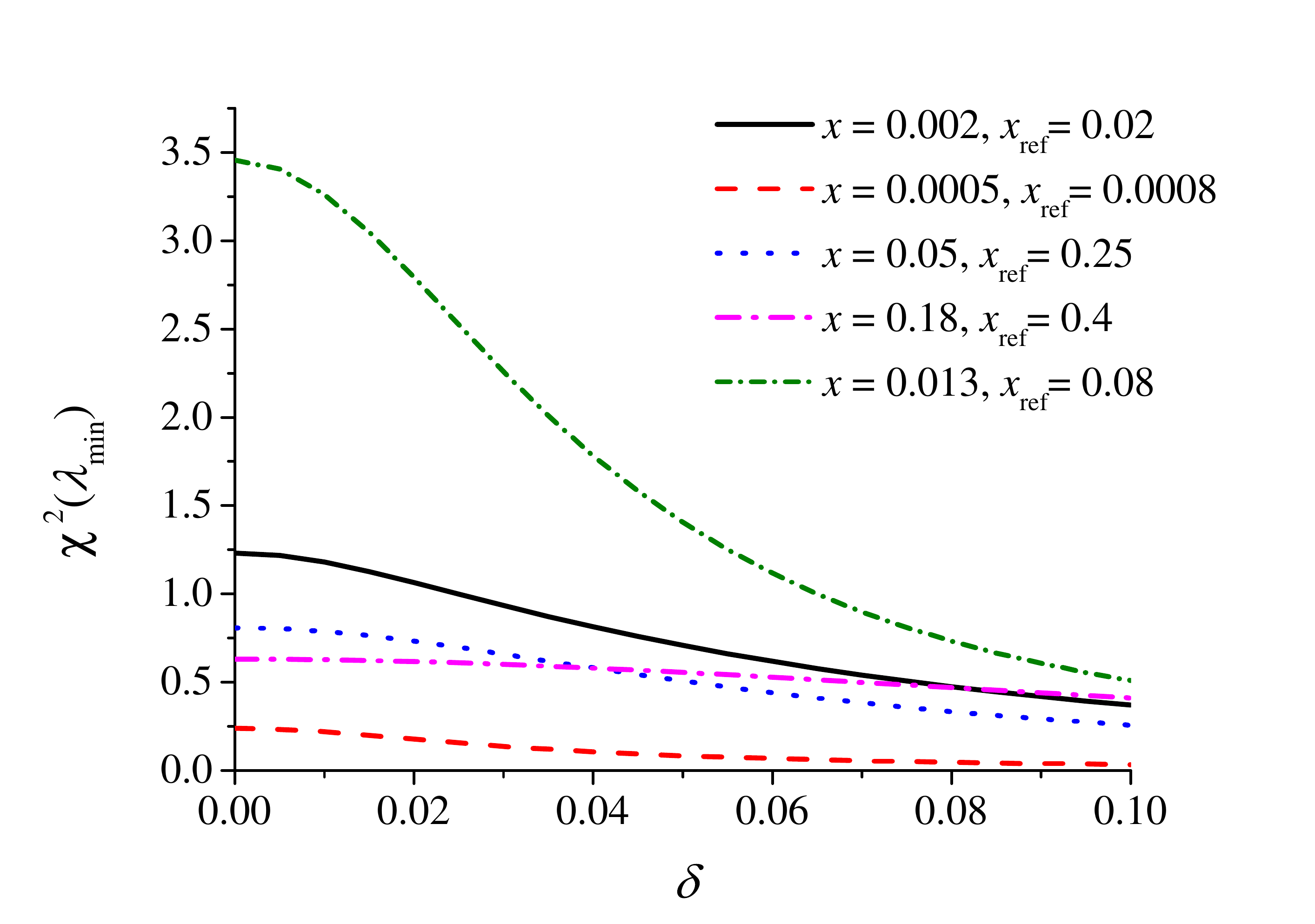}
\caption{Values of $\chi^2(\lambda_{\rm min})$ given by Eq.~(\ref{chix1})
for four typical values of $x_{\rm ref}$ and $x$ used in Fig.~\ref{Rxxref}
and for the case with the largest $\chi^2$ (upper, green curve) as functions of
theoretical error $\delta$.}%
\label{chidel}%
\end{figure}

Before we proceed, let us discuss the influence of  small theoretical
error $\delta$ introduced in Eq.~(\ref{Rxdef}). To this end we plot in
Fig.~\ref{chidel}  $\chi^2(\lambda_{\rm min})$ given by Eq.~(\ref{chix1})
for four typical values of $x_{\rm ref}$ and $x$ used in Fig.~\ref{Rxxref}
and for the case with the highest $\chi^2$  as functions of $\delta$. We
see some moderate decrease of $\chi^2$ which even for $\delta=0$ are in
most cases reasonably small. For the cases with high  $\chi^2$, like
the one represented by the upper curve in
Fig.~\ref{chidel}, the decrease is a bit larger from 3.5 to
2.5 for $\delta=0.03$. 

\begin{figure}[ptb]
\centering
\includegraphics[width=7cm,angle=0]{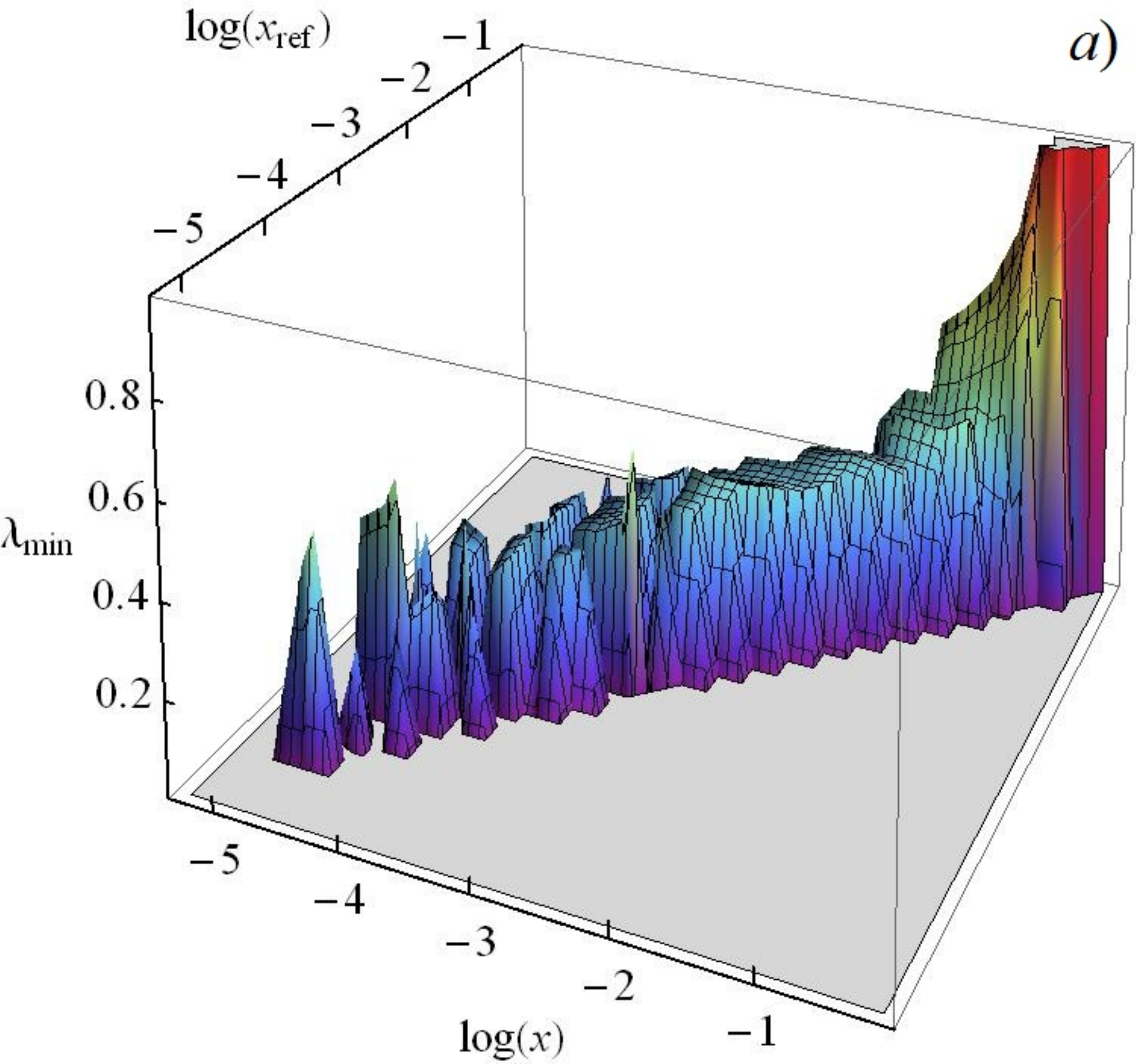}
\includegraphics[width=7.5cm,angle=0]{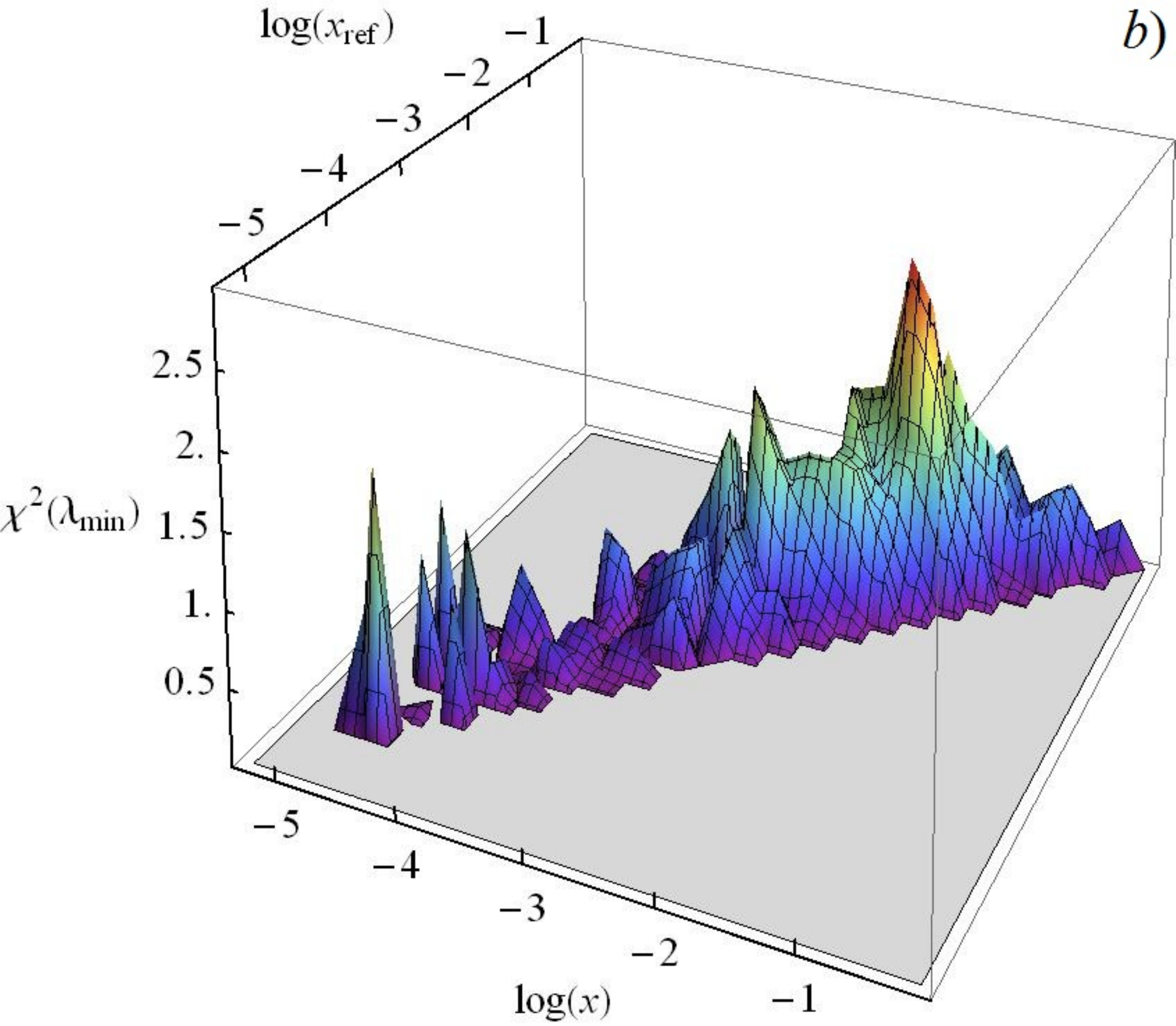}
\caption{Three dimensional
plots of a) $\lambda_{\mathrm{min}}(x,x_{\mathrm{ref}})$
and b) the corresponding $\chi^2(\lambda_{\rm min})$ given by Eq.~(\ref{chix1}).}%
\label{xlamlog}%
\end{figure}

The qualitative measure of geometrical scaling is given by the independence of
$\lambda_{\text{min}}$ on Bjorken $x$ and by the value of $\chi
_{x,x_{\text{ref}}}^{2}(\lambda_{\text{min}})$. In
Fig.~\ref{xlamlog} w show three dimensional  plots of $\lambda_{\text{min}}(x,x_{\text{ref}%
})$ and $\chi_{x,x_{\text{ref}}}^{2}(\lambda_{\text{min}})$ in the
$(x,x_{\text{ref}})$ plane. By construction both $\lambda_{\text{min}}$ and
$\chi^{2}(\lambda_{\text{min}})$ are defined only above the line
$x_{\text{ref}}=x$. We see from Fig.~\ref{xlamlog}.a that the stability corner
of $\lambda_{\text{min}}$ (\emph{i.e.} the region where variations of
$\lambda_{\text{min}}$ are small ) extends up to $x_{\text{ref}}\lesssim0.1$.
In the most of this region $0.3\lesssim\lambda_{\text{min}}\lesssim0.4$.
The value of $\chi_{x,x_{\text{ref}}}^{2}(\lambda_{\text{min}})$
displayed in Fig.~\ref{xlamlog}.b
shows some fluctuations around unity except the region around  
$x_{\rm ref} \sim 0.1$ ({\em i.e.} exactly where $\lambda_{\rm min}$ starts growing)
 where it rises up to $\sim 2.2$.
From Fig.~\ref{xlamlog} we can conclude that
geometrical scaling holds up to Bjorken $x$'s of the order of $10^{-1}$ which
is well above the original expectations.

\begin{figure}[ptb]
\includegraphics[width=7.8cm,angle=0]{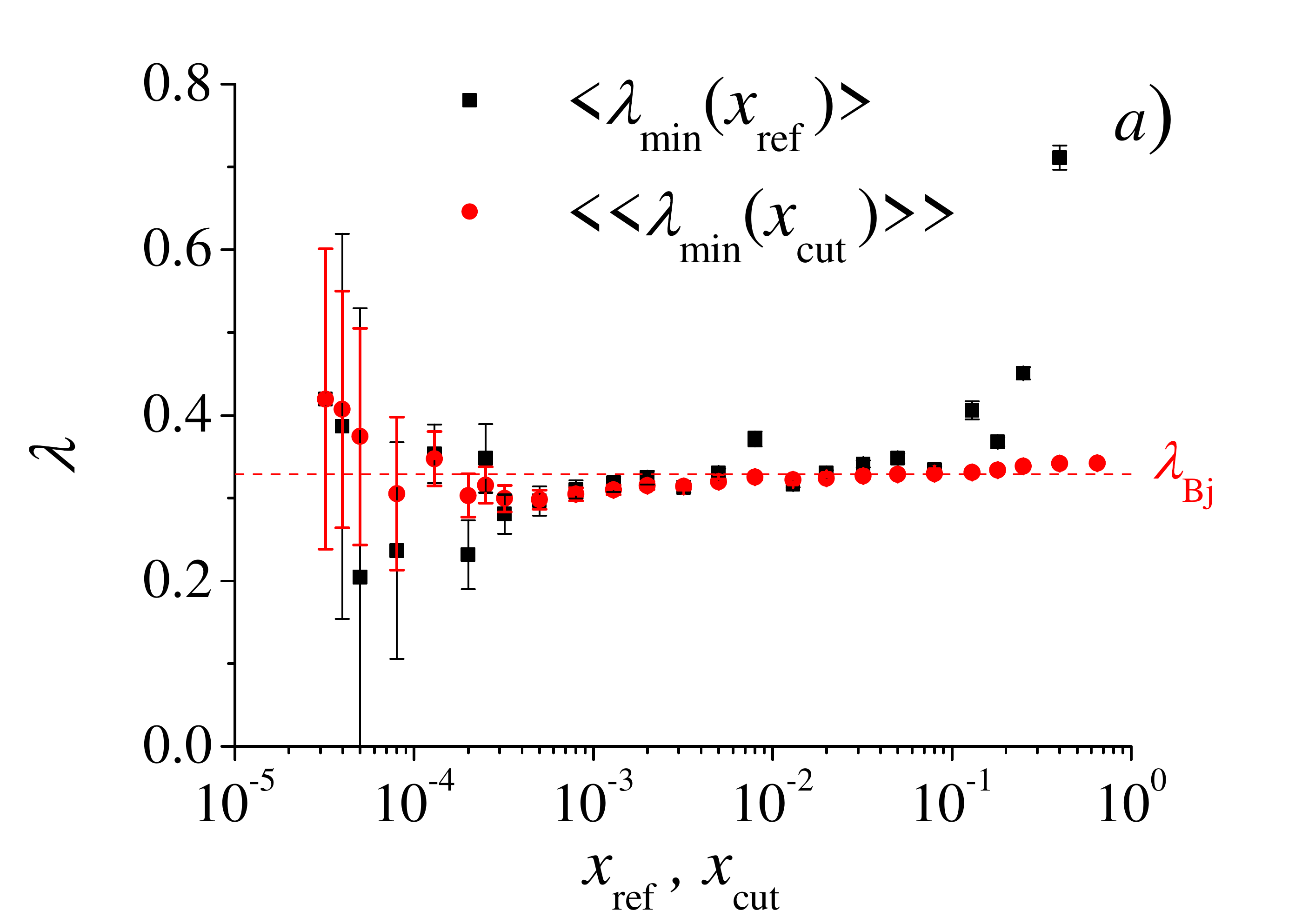}
\includegraphics[width=7.8cm,angle=0]{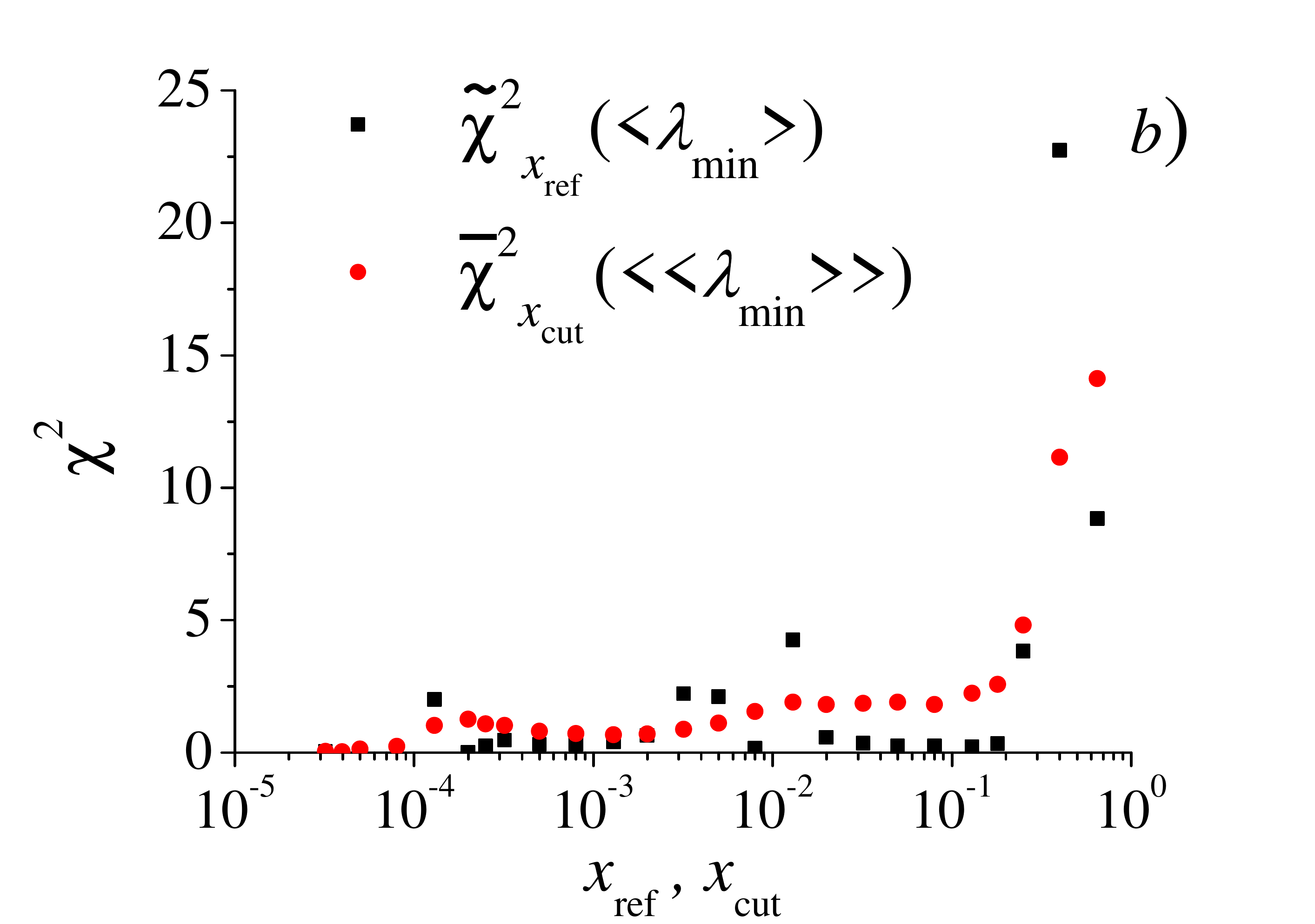}\caption{ Left: averaged
values $\left\langle \lambda_{\mathrm{min}}(x_{\mathrm{ref}})\right\rangle $
(black squares) and $\left\langle \left\langle \lambda_{\mathrm{min}%
}(x_{\mathrm{cut}})\right\rangle \right\rangle $ (red circles); right:
$\tilde{\chi} _{x_{\mathrm{ref}}}^{2}$ (black squares) and $\overline{\chi}
_{x_{\mathrm{cut}}}^{2} $ (red circles) as functions of $x_{\text{ref}}$ and
$x_{\text{cut}}$, respectively.}%
\label{xav}%
\end{figure}

We can now look for possible violation of GS in more quantitative way. In
order to eliminate the dependence of $\lambda_{\text{min}}(x,x_{\mathrm{ref}%
})$ on the value of $x$, we introduce averages over $x$ (denoted in the
following by $\left\langle \ldots\right\rangle $) minimizing the following
chi-square function:
\begin{equation}
\tilde{\chi}_{x_{\text{ref}}}^{2}(\left\langle \lambda\right\rangle )=
\frac{1}{N_{x_{\rm ref}}-1} {\displaystyle\sum\limits_{x<x_{\text{ref}}}} \frac{\left(
\lambda_{\text{min}}(x,x_{\text{ref}})-\left\langle \lambda\right\rangle 
\right)  ^{2}}{\Delta\lambda_{\text{min}}(x,x_{\text{ref}%
})^{2}} \label{chitildex}%
\end{equation}
which gives the "best" value of $\lambda$ denoted as $\langle\lambda
_{\mathrm{min}}(x_{\text{ref}})\rangle$. The sum in (\ref{chitildex}) extends
over all $x$'s such that $\lambda_{\text{min}}(x,x_{\text{ref}})$ exists (see
Fig.~\ref{HERAxref}). $N_{x_{\rm ref}}$ is the number of terms in (\ref{chitildex}). The
results are plotted in Fig. \ref{xav} as black squares. Errors
$\Delta\left\langle \lambda_{\text{min}}\right\rangle $ are calculated from
the requirement that $(N_{x_{\rm ref}}-1)\tilde{\chi}_{x_{\text{ref}}}^{2}%
(\left\langle \lambda\right\rangle )$ changes by 1. For some points one can see very large
errors; this is due to the small number of possible $x$'s for this particular
$x_{\text{ref}}$. On average, 
however, one can approximate $\left\langle
\lambda_{\text{min}}(x_{\text{ref}})\right\rangle $ with a constant value of
$\sim 0.33$ up to $x_{\text{ref}}\sim0.08$ where the rise of 
$\left\langle \lambda_{\text{min}}(x_{\text{ref}})\right\rangle $ should be 
interpreted as a violation of geometrical scaling. The corresponding 
$\tilde{\chi}_{x_{\text{ref}}}^{2}(\langle \lambda_{\mathrm{min}}\rangle)$ 
is reasonably small up to $x_{\rm ref}=0.2$
where it starts rapidly growing.

Looking at Fig.~\ref{xav} one may have an impression that on the average
$\left\langle \lambda_{\text{min}}(x_{\text{ref}})\right\rangle $ is slightly
rising with $x$. This small rise might be, however, attributed to the
dependence of $\lambda$ on $Q^{2}$ through the kinematical correlation between
Bjorken $x$'s and $Q^{2}$ in the kinematical HERA range (see. Fig.
\ref{HERAxref}). $Q^{2}$ dependence of $\lambda$ has been measured in DIS for
very small $x$'s \cite{HERAdata} and has been discussed in the context of the
$p_{\mathrm{T}}$ spectra at the LHC \cite{Praszalowicz:2011tc}. It is
theoretically motivated by the corrections due the DGLAP evolution
\cite{Bartels:2002cj,Kowalski:2010ue}. We shall come back to this possibility
in Sect.~\ref{sumout}.

To quantify further the hypothesis of geometrical scaling with constant
$\lambda$ we form yet another chi-square function%
\begin{equation}
\overline{\chi}_{x_{\text{cut}}}^{2}(\left\langle \left\langle \lambda
\right\rangle \right\rangle )=\frac{1}{N_{x_{\text{cut}}}-1}%
{\displaystyle\sum\limits_{x_{\text{ref}} \leq x_{\text{cut}}}}\,
{\displaystyle\sum\limits_{x<x_{\text{ref}}}} \frac{\left(  \lambda
_{\text{min}}(x,x_{\text{ref}})-\left\langle \left\langle \lambda
\right\rangle \right\rangle \right)  ^{2}}{\Delta
\lambda_{\text{min}}(x,x_{\text{ref}})^{2}} \label{titilchi2}%
\end{equation}
which we minimize to obtain $\left\langle \left\langle \lambda_{\text{min}%
}(x_{\text{cut}})\right\rangle \right\rangle $.

The idea behind equation (\ref{titilchi2}) is to see how well one can fit
$\left\langle \lambda_{\text{min}}(x_{\text{ref}})\right\rangle $ with a
constant $\lambda$ up to $x_{\text{ref}}=x_{\text{cut}}$. Were there any
strong violations of GS above some $x_{0}$, one should see a rise of
$\left\langle \left\langle \lambda_{\text{min}}(x_{\text{cut}})\right\rangle
\right\rangle $ once $x_{\text{cut}}$ becomes larger than $x_{0}$. As can be
seen from Fig.~\ref{xav}, where $\left\langle \left\langle \lambda
_{\text{min}}(x_{\text{cut}})\right\rangle \right\rangle $ and the
corresponding $\overline{\chi}_{x_{\text{cut}}}^{2}$ are plotted as red
circles, no drastic change in $\left\langle \left\langle \lambda_{\text{min}%
}(x_{\text{cut}})\right\rangle \right\rangle $ can be seen, although the slow
rise is seen above $x_{\text{cut}}\simeq0.1$. On the contrary
$\overline{\chi}_{x_{\text{cut}}}^{2}$ starts to 
rise slowly at $x_{\rm cut}=0.08$ and then rises steeply above $x_{\text{cut}%
}\simeq0.2$.  

Summarizing discussion of Fig.~\ref{xav} we conclude
that the best value of a constant $\lambda$ corresponds to
$x_{\text{cut}}=0.08$ which we denote as
\begin{equation}
\lambda_{\text{Bj}}=0.329\pm0.002 \label{lBjfinal}%
\end{equation}
where subscript "Bj" stands for Bjorken $x$ binning. The error is purey
statistical, we shall discuss systematic uncertainties in the end
of Sect.~\ref{Wbinning} and in Sect.~\ref{sumout}.

Let us finish by a remark on comparison of $\left\langle \left\langle
\lambda_{\text{min}}(x_{\text{cut}})\right\rangle \right\rangle $ with a more
"differential" quantity $\left\langle \lambda_{\text{min}}(x_{\text{ref}%
})\right\rangle $ depicted in Fig.~\ref{xav}.a.
Since $\left\langle \left\langle \lambda_{\text{min}%
}(x_{\text{cut}})\right\rangle \right\rangle $ is in a sense an average of all
$\left\langle \lambda_{\text{min}}(x_{\text{ref}})\right\rangle $ for
$x_{\text{ref}}\le x_{\text{cut}}$, the rapid increase of $\left\langle
\lambda_{\text{min}}(x_{\text{ref}})\right\rangle $ above $x_{\text{ref}%
}\simeq0.08$ is smoothed out due to a long constant tail of $\left\langle
\lambda_{\text{min}}(x_{\text{ref}})\right\rangle $ for smaller $x_{\text{ref}%
}$'s.  Below $x=0.08$ we have that $\left\langle
\lambda_{\text{min}}(x)\right\rangle \simeq\left\langle \left\langle
\lambda_{\text{min}}(x)\right\rangle \right\rangle $ which further confirms
validity of geometrical scaling in this region.

\section{Energy binning}

\label{Wbinning}

\begin{figure}[t]
\centering
\includegraphics[width=10cm]{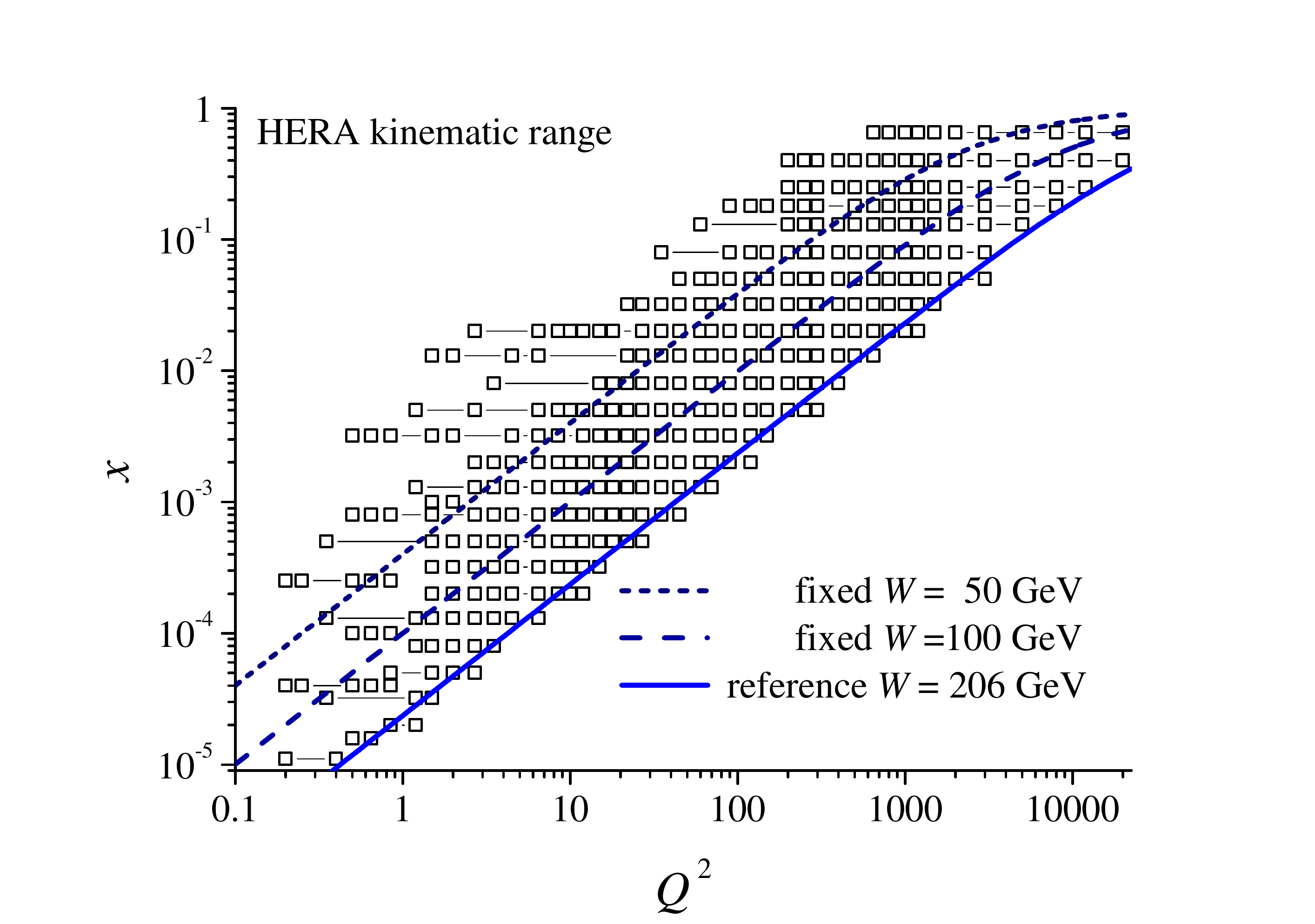}\caption{HERA kinematic range in
$(Q^{2},x)$ plane with several lines of constant $W$.}%
\label{HERAWref}%
\end{figure}

Here we are going to repeat analysis of Sect.~\ref{Bjxbinning} for the
combined HERA data \cite{HERAcombined}, but in bins of $(W,Q^{2})$ rather than
$(x,Q^{2})$. The reason is that fixed energy points span over much wider
common $Q^{2}$ range (see Fig. \ref{HERAWref}) and therefore we do not have
to choose different reference points and average over them. Moreover, analysis
in terms of $W$ follows exactly the method used to analyze $p_{\text{T}}$
spectra at the LHC \cite{McLerran:2010ex,Praszconf}. The disadvantage consists
in the fact that HERA points are not organized in the energy bins, therefore
we have to "rebin" them using formula (\ref{xdef}). This leads to the loss of
statistics. Indeed, instead of 432 points (see Sect. \ref{Bjxbinning}) we
shall use here 347 $e^{+}p$ data points. Furthermore some ambiguities arise,
such as the choice of the bin size, the need to recalculate Bjorken $x$ and
others discussed in detail in Ref.~\cite{Stebel:2012ky}. In what follows we
shall use logarithmic binning with step 1.3 -- every consecutive $W$ is 1.3
times larger than preceding one (in Ref.~\cite{Stebel:2012ky} different
binnings have been also considered without any major change of the results). 
We define value of energy $W$ as the mean of two limiting values
$W_{\mathrm{min}}^{\prime}$ and $W_{\mathrm{max}}^{\prime}$ between which it
lies (see Table \ref{table1}).

\begin{table}[ptb]
\begin{center}%
\begin{tabular}
[c]{|c|c|c|c|c|c|c|c|}\hline
$W^{\prime}_{\mathrm{min}}\mathrm{\ [GeV]}$ & 10 & 13 & 16.9 & 22 & 28.6 &
37.1 & 48.3\\\hline
$W^{\prime}_{\mathrm{max}}\mathrm{\ [GeV]}$ & 13 & 16.9 & 22 & 28.6 & 37.1 &
48.3 & 62.7\\\hline
$W\mathrm{\ [GeV]}$ & 11.5 & 15 & 19.4 & 25.3 & 32.8 & 42.7 & 55.5\\\hline
Number of points & 6 & 3 & 13 & 22 & 22 & 32 & 33\\\hline\hline
$W^{\prime}_{\mathrm{min}}\mathrm{\ [GeV]}$ & 62.7 & 81.6 & 106 & 137.9 &
179.2 & 233 & \\\hline
$W^{\prime}_{\mathrm{max}}\mathrm{\ [GeV]}$ & 81.6 & 106 & 137.9 & 179.2 &
233 & 302.9 & \\\hline
$W\mathrm{\ [GeV]}$ & 72.2 & 93.8 & 122 & 158.5 & 206.1 & 267.9 & \\\hline
Number of points & 40 & 40 & 42 & 43 & 44 & 7 & \\\hline
\end{tabular}
\end{center}
\caption{Energy bins and energies assigned to them. Number of points in
different bins is also displayed (these are values for $e^{+}p$ data).}%
\label{table1}%
\end{table}

In what follows we use $W_{\mathrm{ref}}=206$ GeV because it gives the widest
range of $\tau$ values and is one of the biggest energies that we have at our
disposal (GS is expected to be present at large energies), and it has the
largest number of points in $Q^{2}$. In Ref.~\cite{Stebel:2012ky} different
choices of $W_{\mathrm{ref}}$ have been also analyzed with essentially the
same conclusions.

Similarly to the case of $x$--binning we shall construct ratios of
$\gamma^{\ast}p$ cross-sections at different $W$'s as functions of $\tau$. In
general $\tau$ values for $W_{\mathrm{ref}}$ and $W$ are different so we need
to interpolate the reference cross-section to the value of $Q_{\text{ref}}%
^{2}$ which corresponds to the $\tau$ value needed to calculate the ratio.
Here we follow closely the method of Sect.\ref{Bjxbinning} using linear
interpolation in $\log Q^{2}$.

Now, for every point with energy $W_{i}$ we define a ratio 
($k$ labels points with energy $W_{i}$ but of different $Q^{2}$):
\begin{equation}
R_{W_{i},W_{\text{ref}}}(\lambda;\tau_{k}):=\frac{\sigma_{\gamma^{\ast}p}%
(W_{\text{ref}},\tau(W_{\text{ref}},Q^2_{k,\text{ref}};\lambda))}{\sigma_{\gamma^{\ast}p}%
(W_i,\tau(W_{i},Q^2_{k};\lambda))}\;\text{with}\;\tau_{k}=\tau(W_{i}%
,Q_{k}^{2};\lambda)=\tau(W_{\text{ref}},Q_{k,\text{ref}}^{2};\lambda).
\label{ratiodef}%
\end{equation}
Uncertainty of ratio $R_{W_{i},W_{\text{ref}}}(\lambda;\tau_{k})$ is given
by:
\begin{align}
& \Delta R_{W_{i},W_{\text{ref}}}(\lambda;\tau_{k})^{2} =   \label{errW} \\
& \left( \left( 
\frac{\Delta\sigma_{\gamma^{\ast}p}(W_{i},\tau(W_{i},Q_{k}^{2}))}{
\sigma_{\gamma^{\ast}p}(W_{i},\tau(W_{i},Q_{k}^{2}))}\right)^{2} 
+\left(  \frac{\Delta\sigma_{\gamma^{\ast}p}(W_{\text{ref}},\tau(W_{\text{ref}},Q_{k,\text{ref}}%
^{2}))}%
{\sigma_{\gamma^{\ast}p}(W_{\text{ref}},\tau(W_{\text{ref}},Q_{k,\text{ref}}^{2}))}%
\right)  ^{2}\right) R_{W_{i},W_{\text{ref}}}(\lambda; \tau_k)^2 
+ \delta^2 & \nonumber %
\end{align}
Since we have fixed $W_{\text{ref}}=206$ GeV, in what follows we shall omit
subscript $W_{\text{ref.}}$

\begin{figure}[ptb]
\centering
\includegraphics[width=10cm,angle=0]{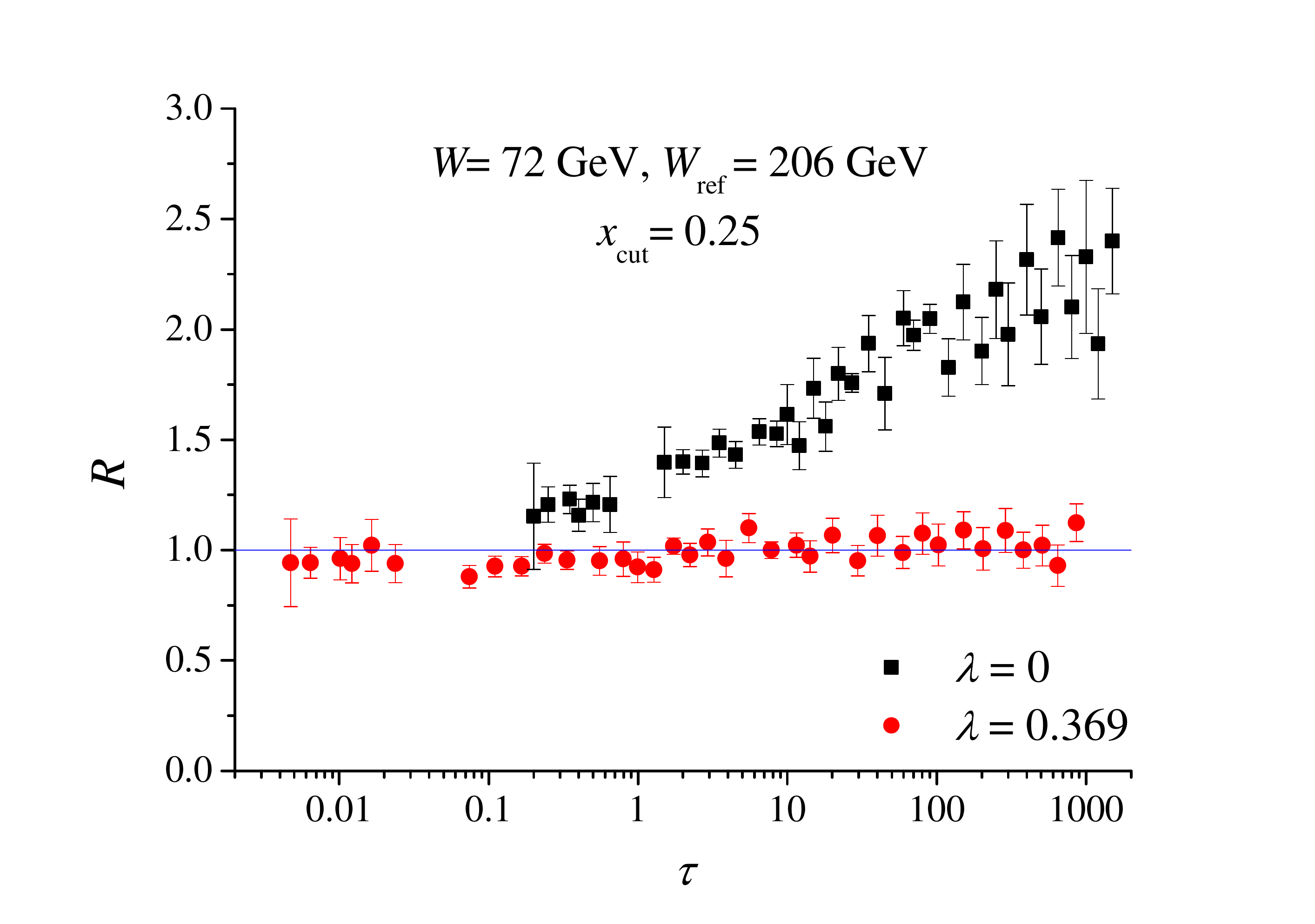}\caption{Ratios $R_{72}(\lambda)$
for $\lambda=0$ and $\lambda=\lambda_{\mathrm{min}}=0.369$ both with
$x_{\mathrm{cut}}=0.25$. }%
\label{zesWyk4}%
\end{figure}

In Fig.~\ref{zesWyk4} we show as an example $R_{72}$ plotted for $\lambda=0$
and $\lambda=0.369$. We see that $R_{72}$ decreases several times when we use
scaling variable rather than $Q^{2}$. This is generic feature which we employ
to look for geometrical scaling.

Our aim is to find such $\lambda$ for given energy $W_{i}$
that deviations $R_{W_{i}}(\lambda;\tau_{k})-1$ are minimal. Taking into
account uncertainties $\Delta R_{W}$ we define the chi-square function:%

\begin{equation}
\chi_{W_{i}}^{2}(x_{\mathrm{cut}};\lambda)=\frac{1}{N_{W_{i},x_{\text{cut}}%
}-1}\sum\limits_{k\in W_{i};\,x\leq x_{\mathrm{cut}}}\frac{(R_{W_{i}}%
(\lambda;\tau_{k})-1)^{2}}{\Delta R_{W_{i}}(\lambda;\tau_{k})^{2}}
\label{defchi2}%
\end{equation}
where $k\in W_{i};$ $x\leq x_{\mathrm{cut}}$ means that we sum over points
corresponding to given energy $W_{i}$ and values of $x$ that are not larger
than $x_{\mathrm{cut}}$. The reason to introduce a cut-off on Bjorken $x$ is
to look for violations of GS once we get into the region of large $x$'s.

\begin{figure}[ptb]
\centering
\includegraphics[width=4.5cm,angle=0]{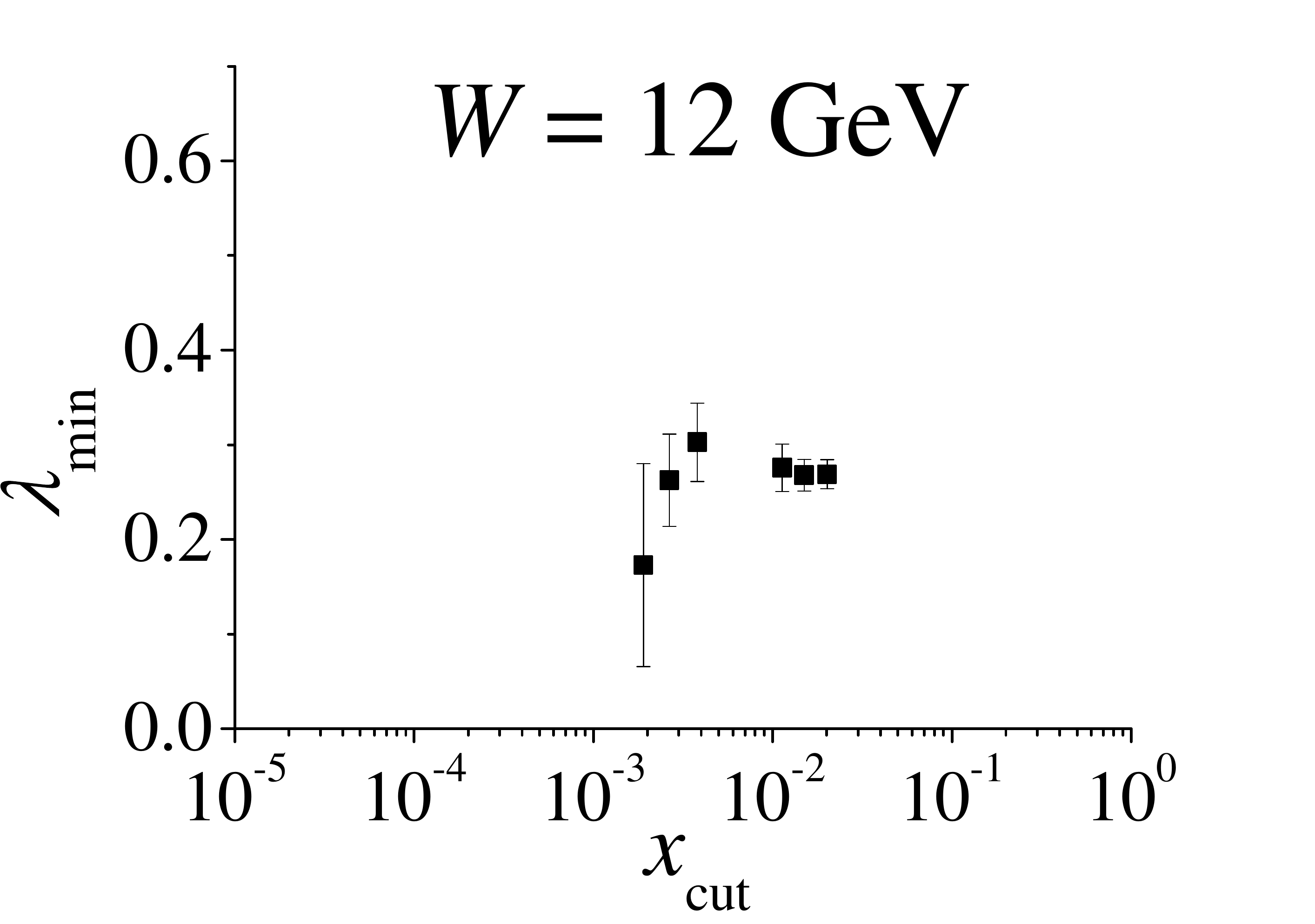}
\includegraphics[width=4.5cm,angle=0]{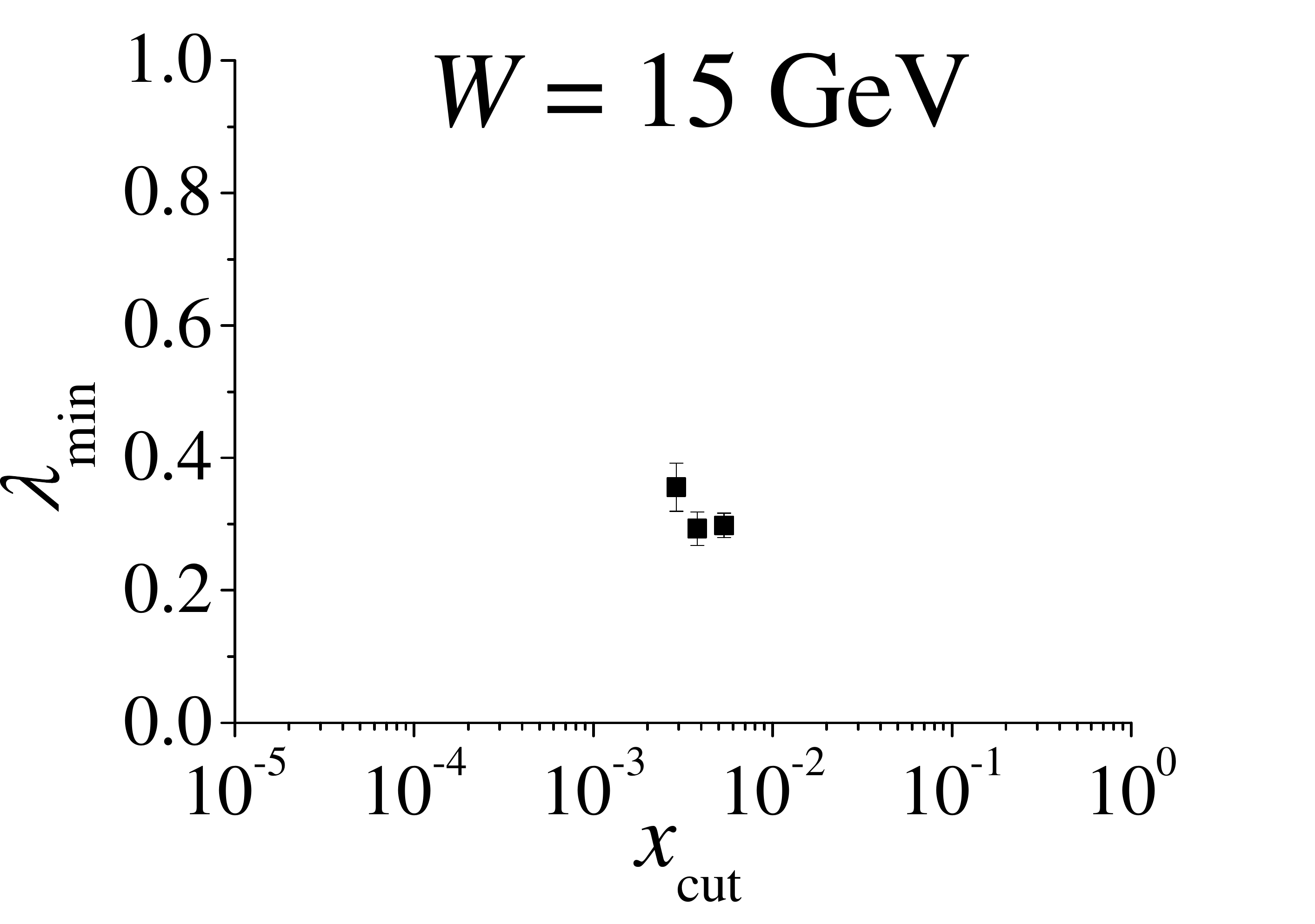}
\includegraphics[width=4.5cm,angle=0]{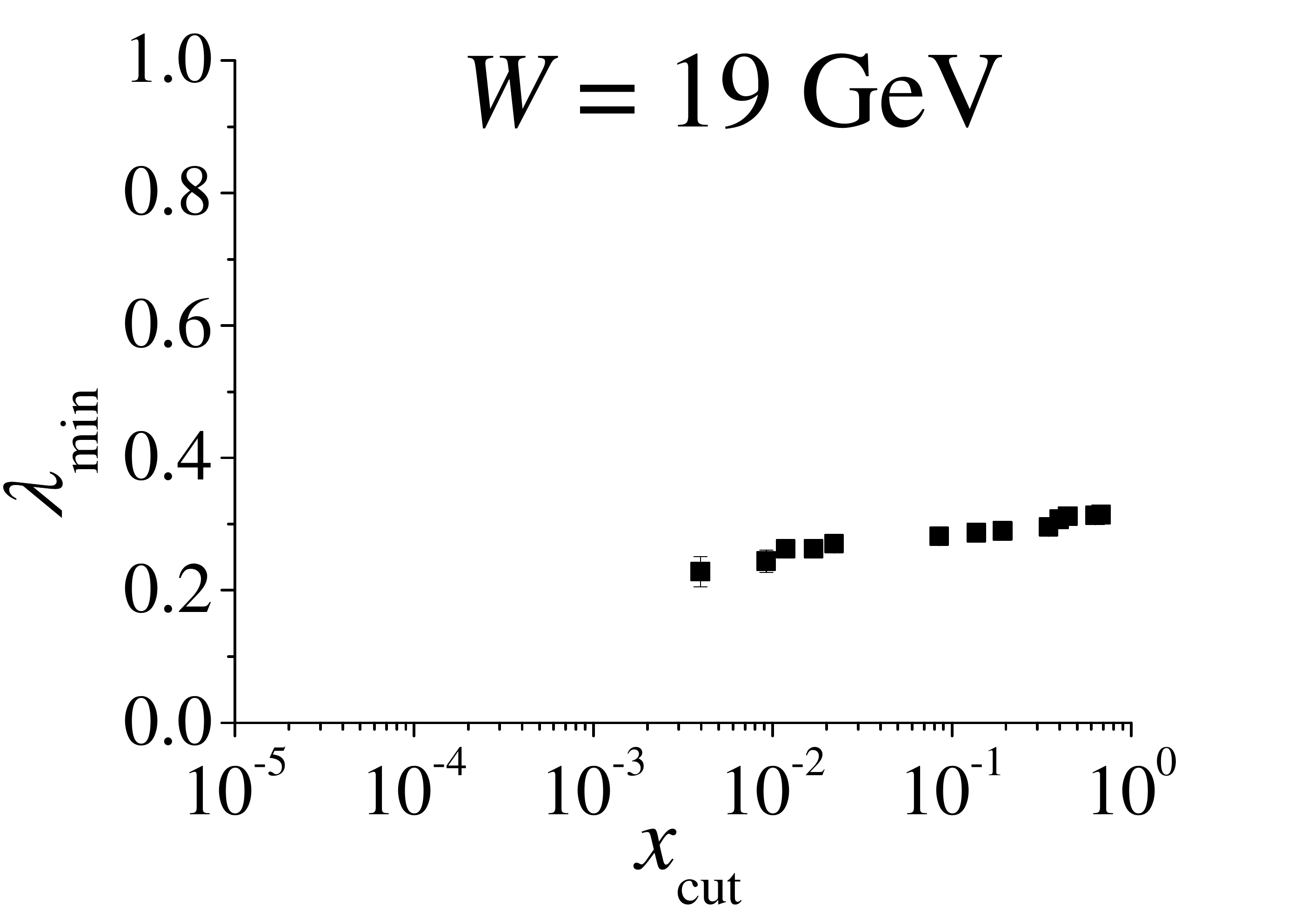}\newline%
\includegraphics[width=4.5cm,angle=0]{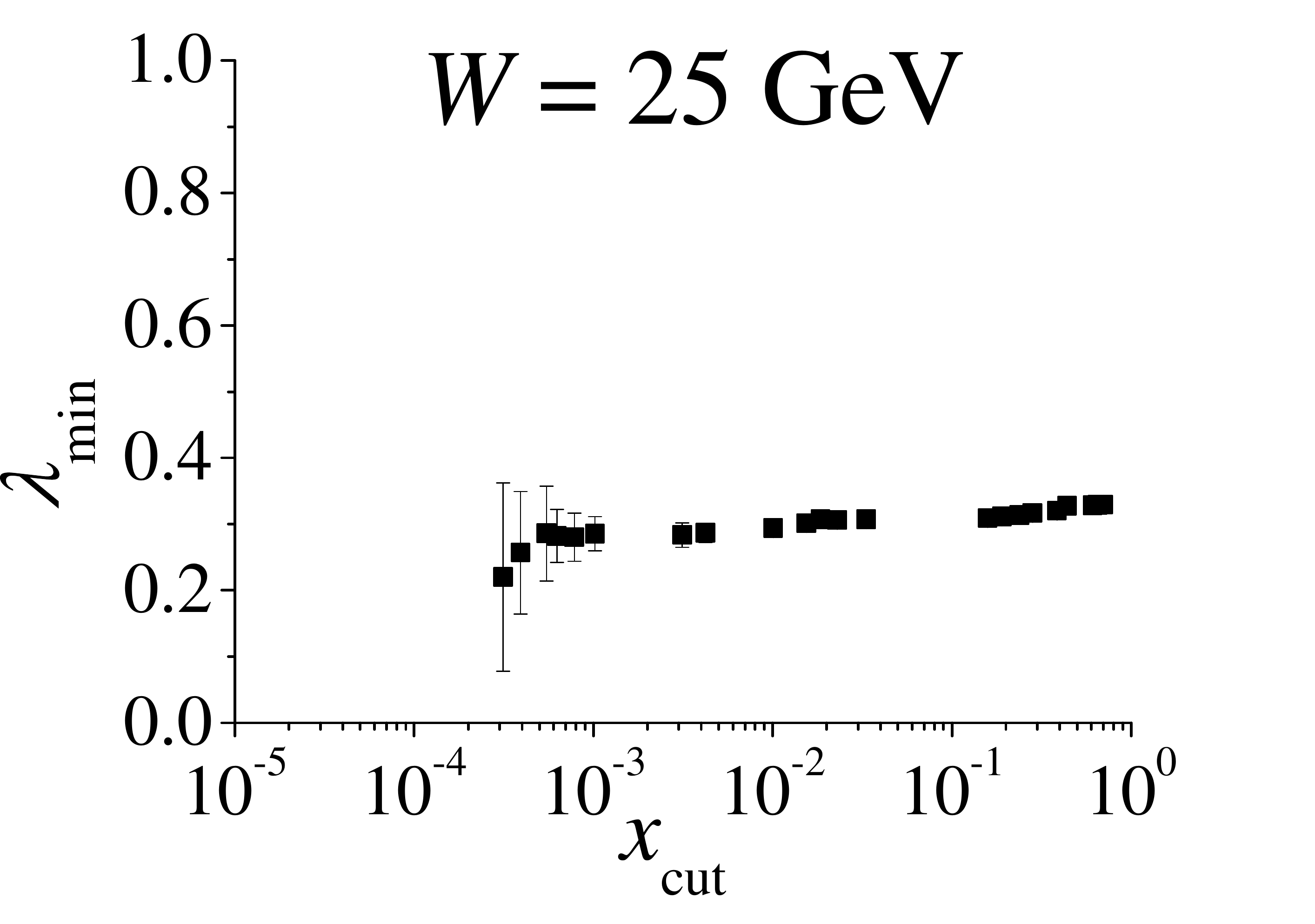}
\includegraphics[width=4.5cm,angle=0]{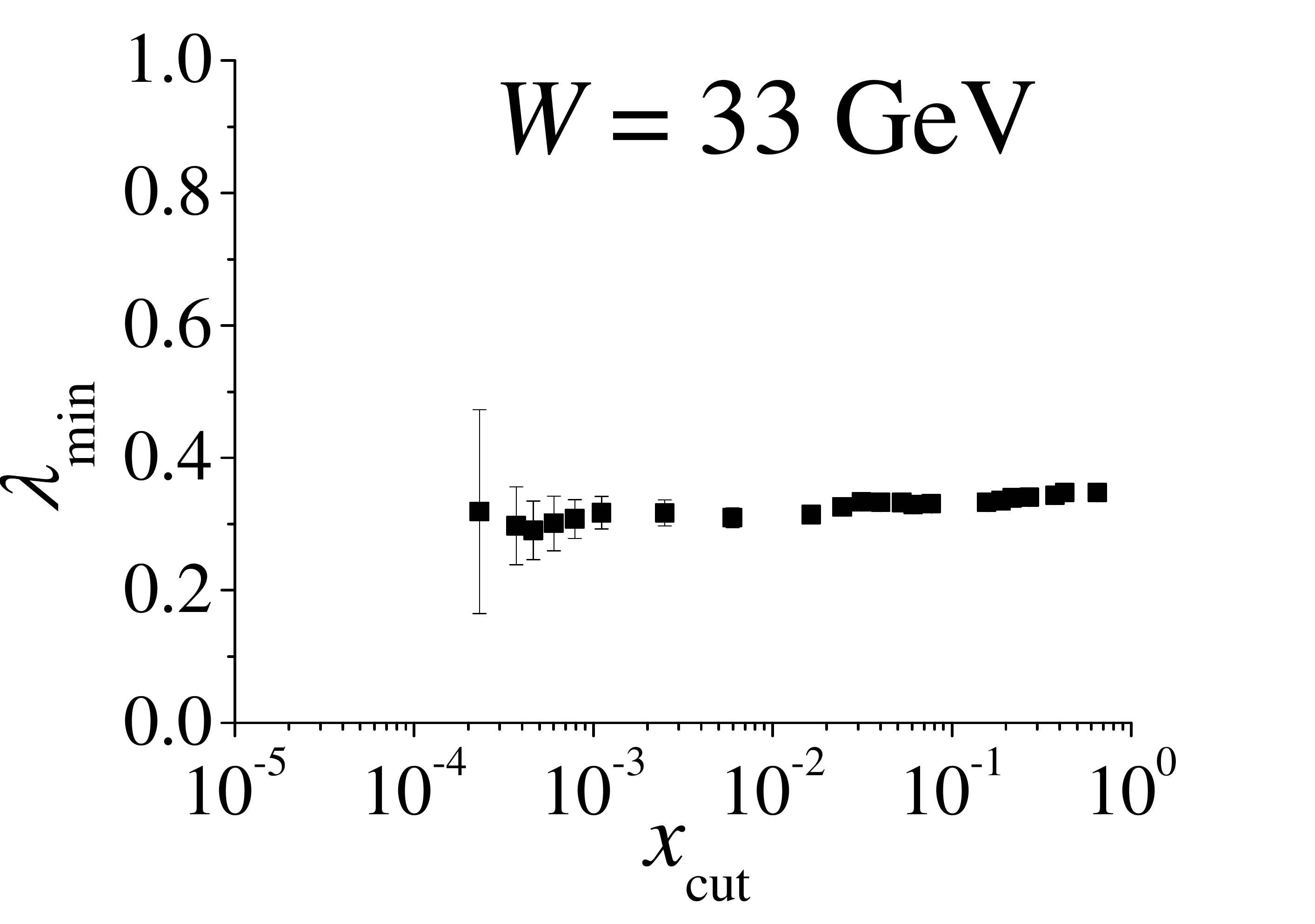}
\includegraphics[width=4.5cm,angle=0]{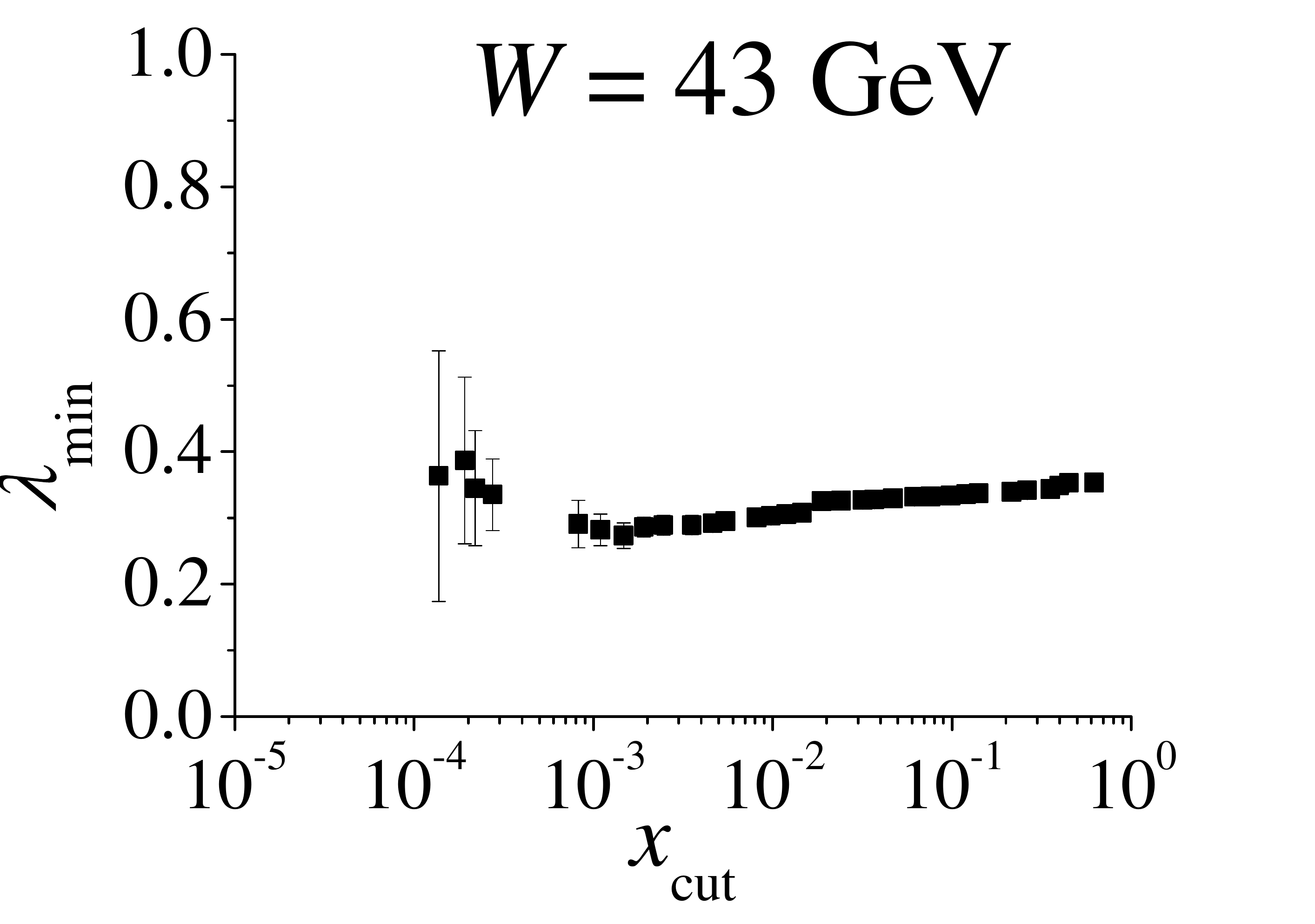}\newline%
\includegraphics[width=4.5cm,angle=0]{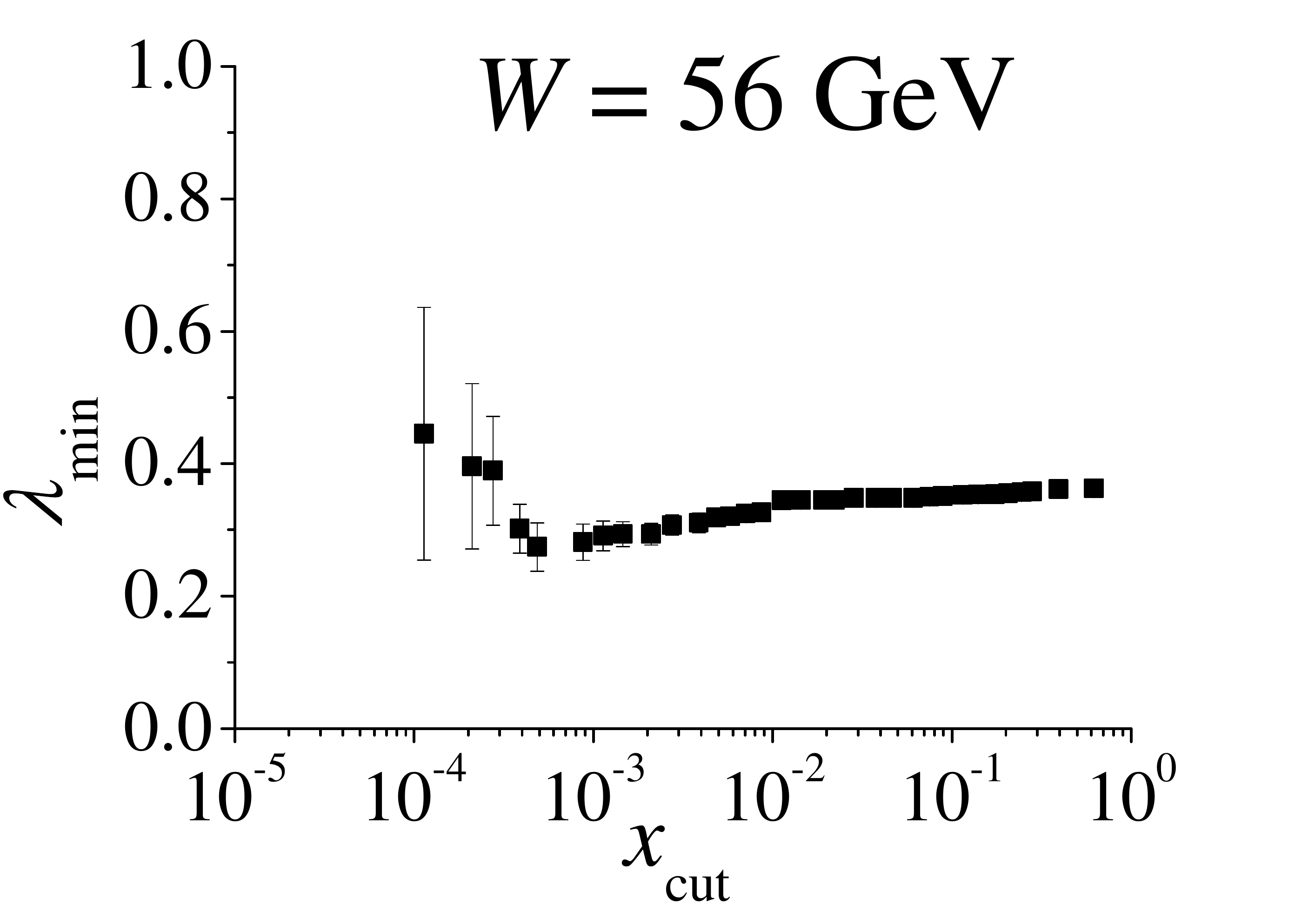}
\includegraphics[width=4.5cm,angle=0]{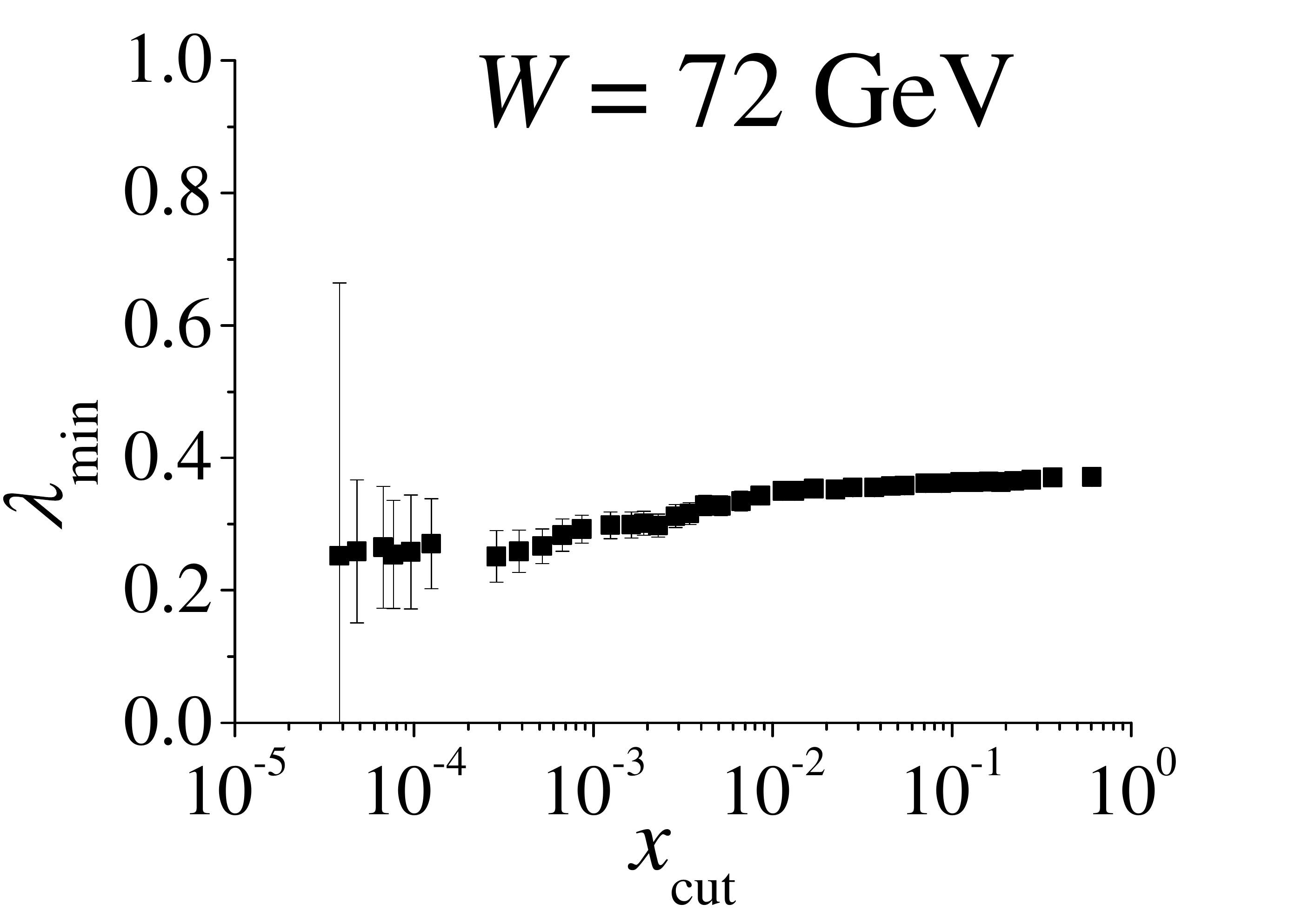}
\includegraphics[width=4.5cm,angle=0]{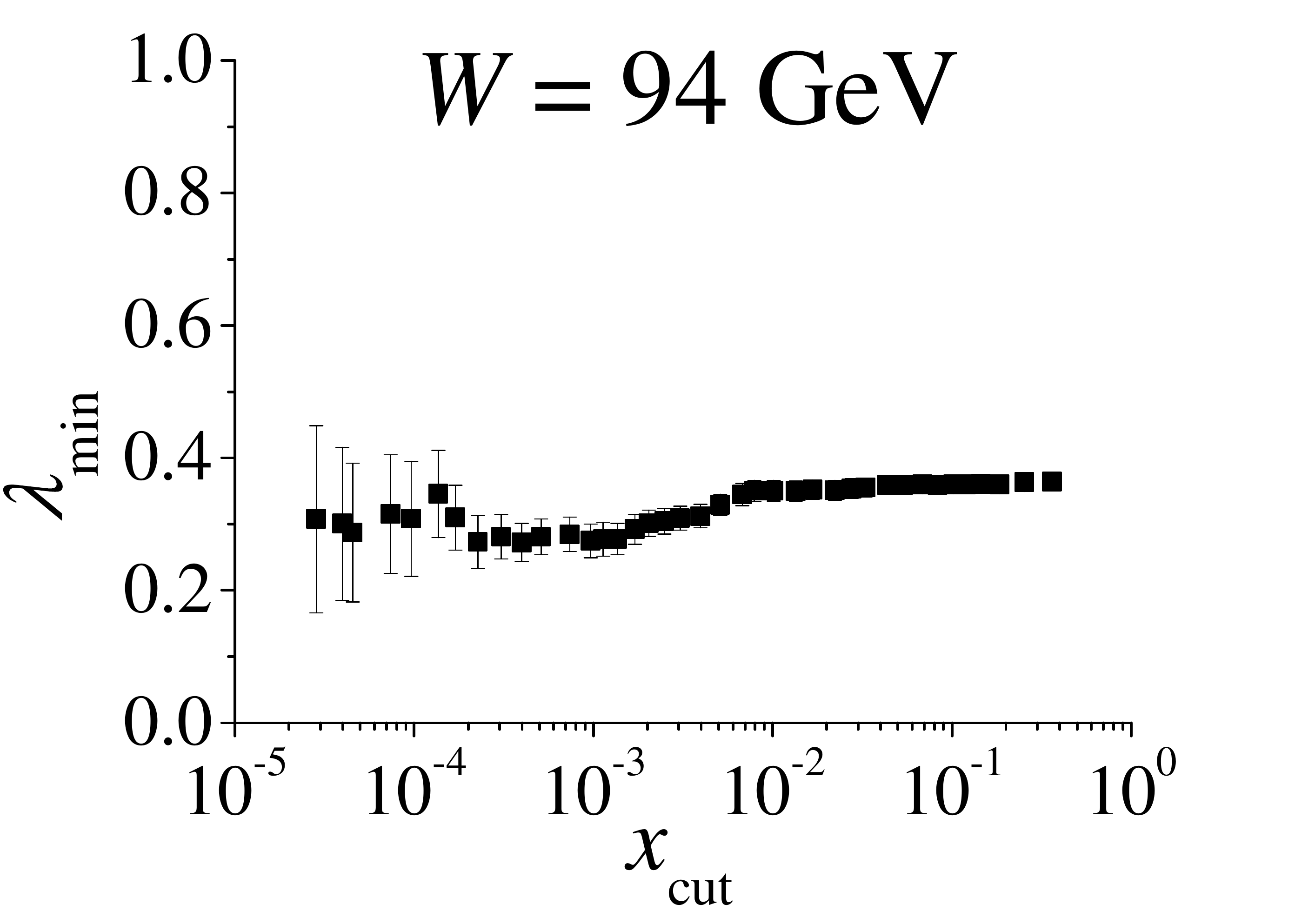}\newline%
\includegraphics[width=4.5cm,angle=0]{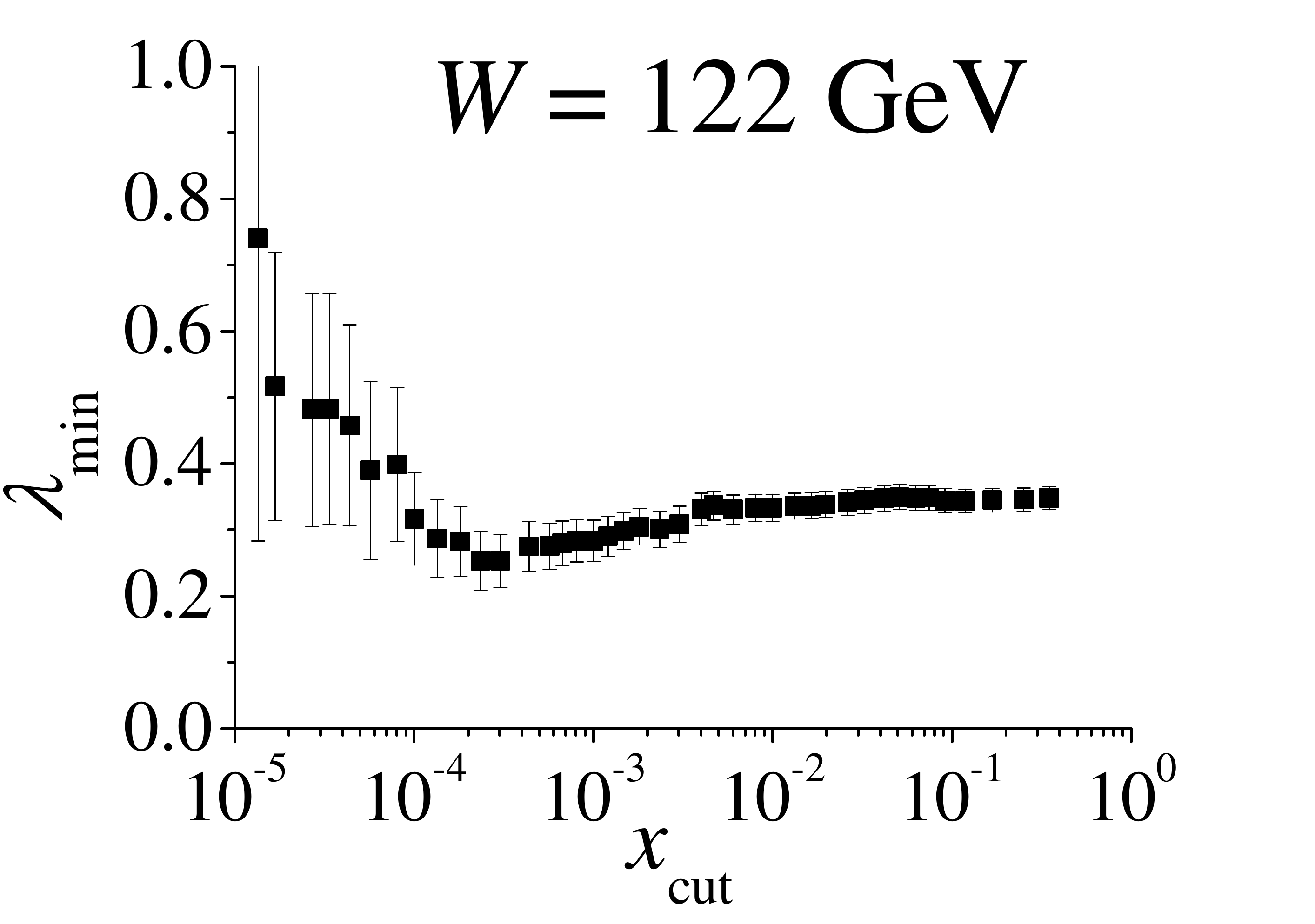}
\includegraphics[width=4.5cm,angle=0]{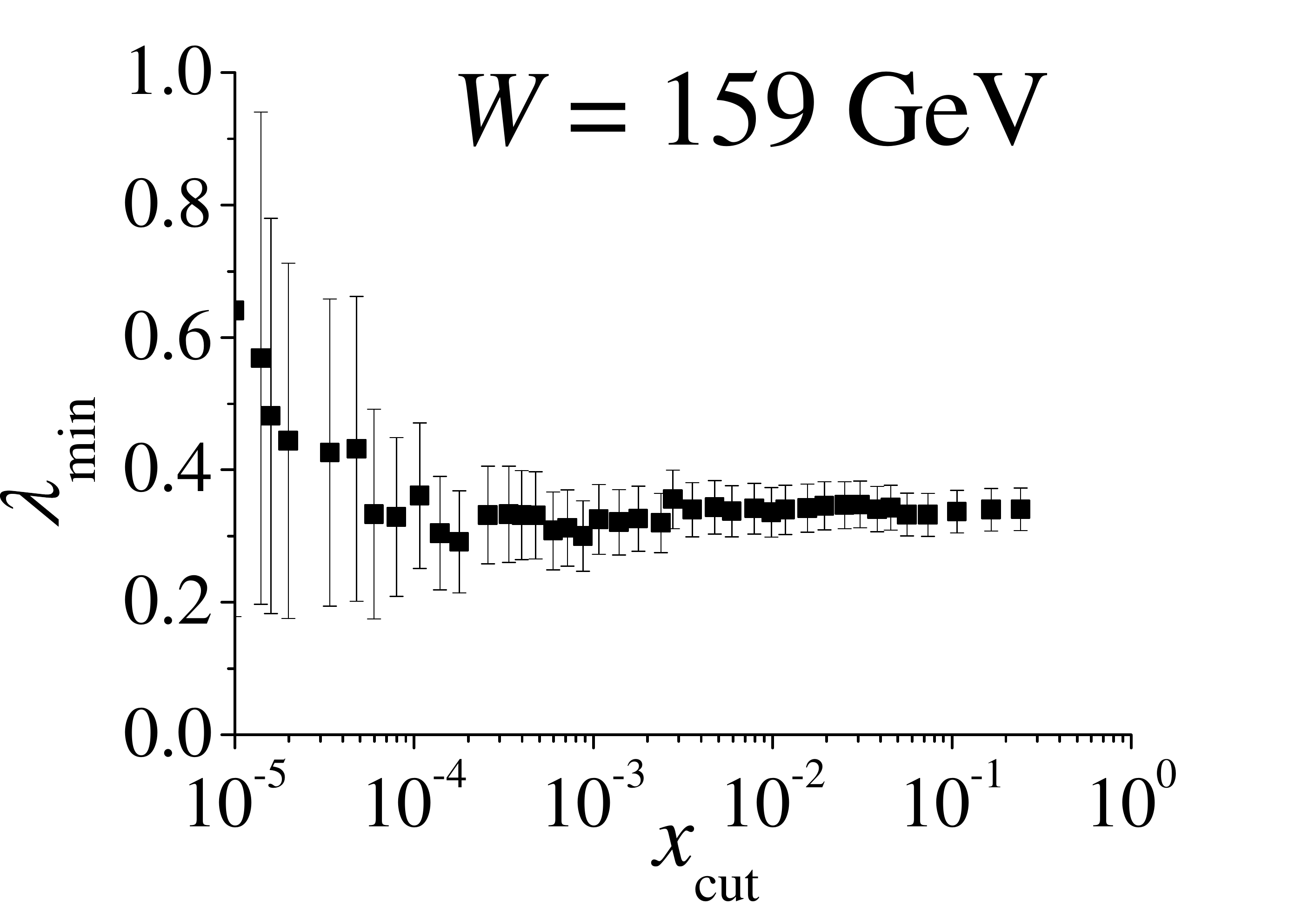}
\includegraphics[width=4.5cm,angle=0]{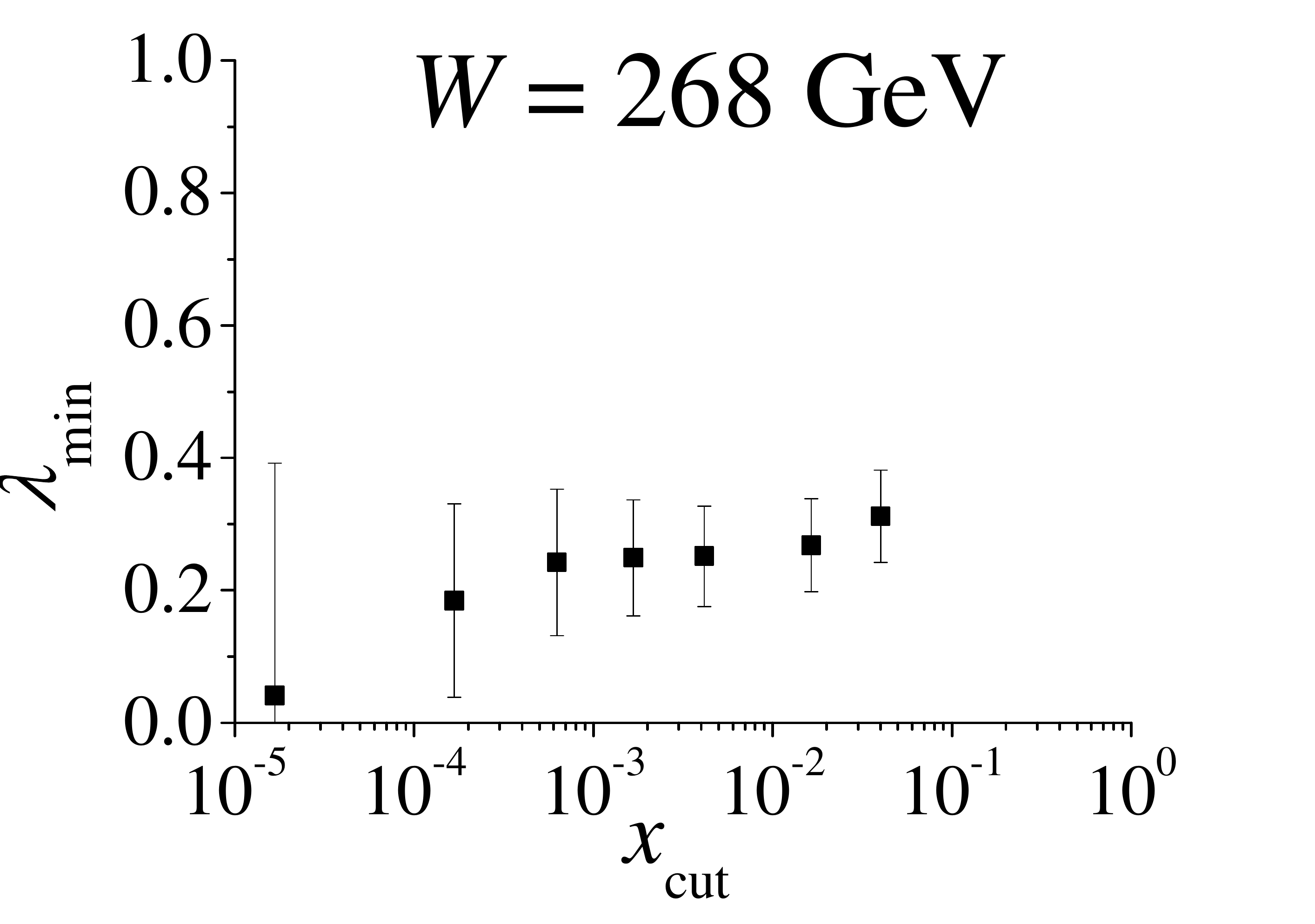}\newline\caption{Exponent
$\lambda_{\mathrm{min}}$ as a function of $x_{\mathrm{cut}}$ for all energies
$W\neq W_{\mathrm{ref}}=206$ GeV.}%
\label{zesWyk5}%
\end{figure}

\begin{figure}[ptb]
\centering
\includegraphics[width=7.5cm,angle=0]{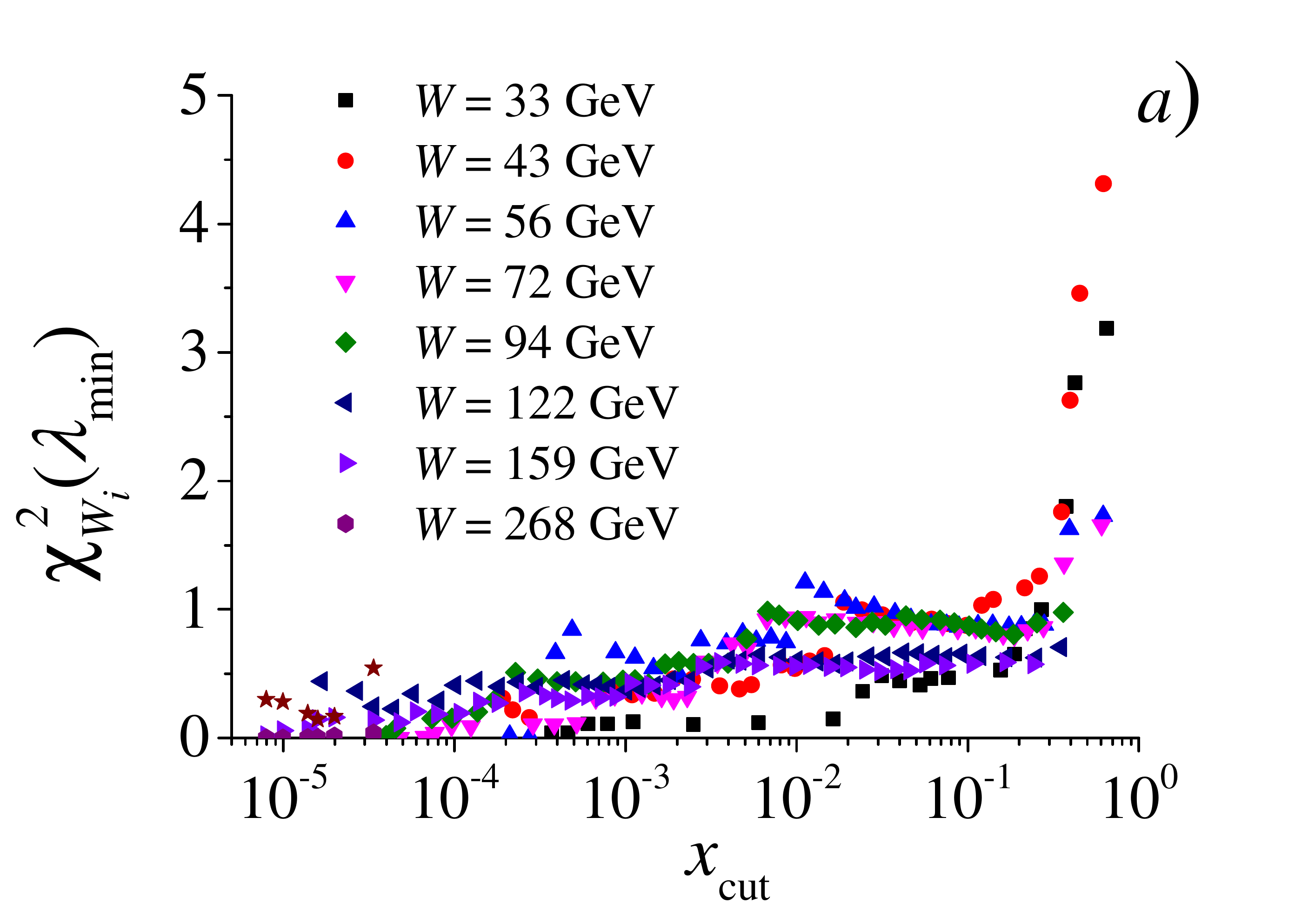}
\includegraphics[width=7.5cm,angle=0]{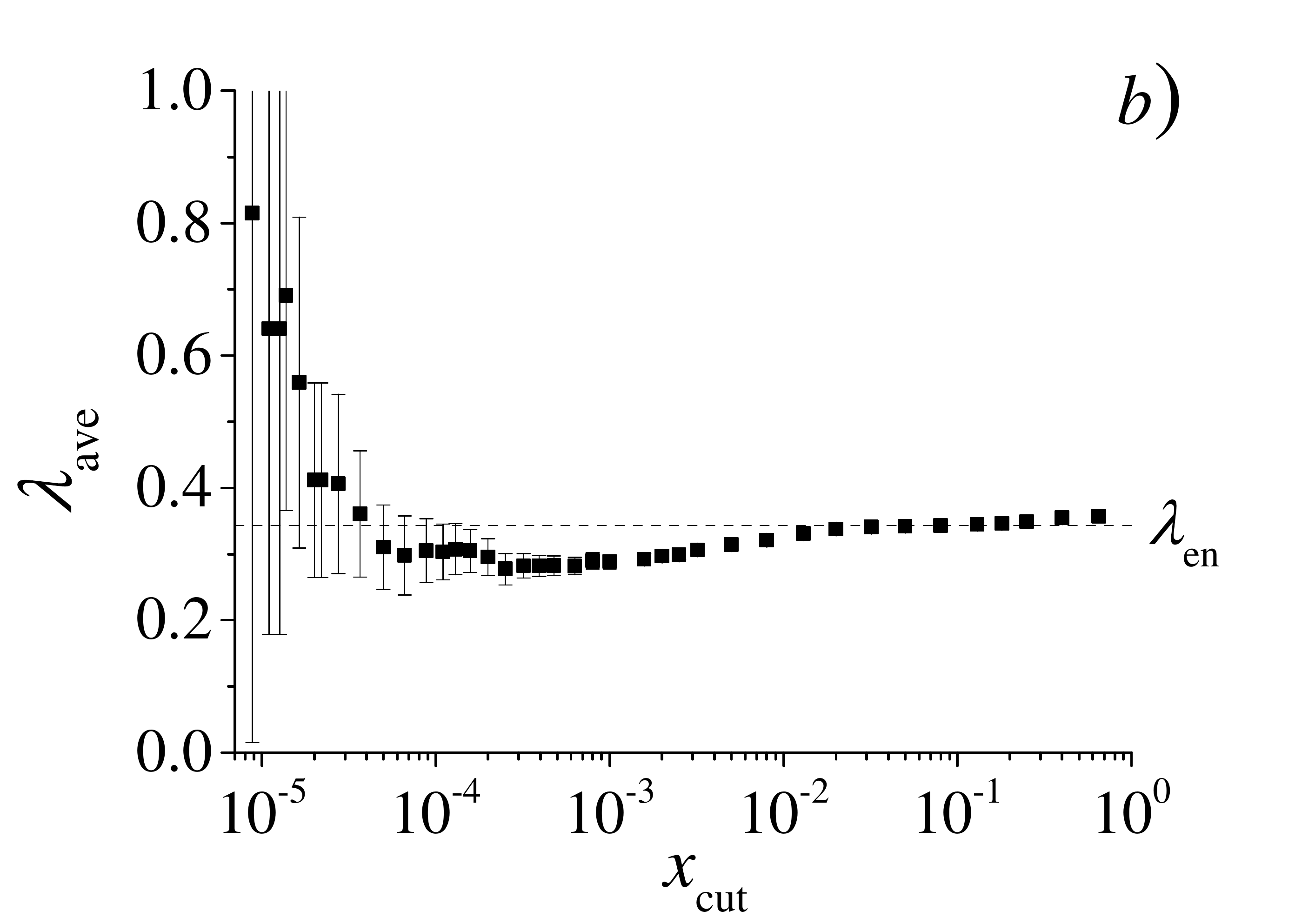}
\caption{Left panel: plot of
$\chi_{W_{i}}^{2}$ for different energies as functions of $x_{\text{cut}}$.
Right panel: exponent $\lambda_{\mathrm{min}}$ averaged over energies (denoted as $\lambda_{\mathrm{ave}}$) as a function of $x_{\mathrm{cut}}$ .}%
\label{Wchixcut}%
\end{figure}

We will search $\lambda_{\mathrm{min}}\left(  W_{i},x_{\mathrm{cut}}\right)  $
which minimizes $\chi_{W_{i}}^{2}$ for given $W_{i}$ and $x_{\mathrm{cut}}$.
Uncertainty of $\lambda_{\mathrm{min}}$ is estimated by requiring that
$(N_{W_{i},x_{\text{cut}}}-1)\chi_{W_{i}}^{2}(x_{\mathrm{cut}};\lambda)$
changes by 1 when $\lambda$ is varied around the minimum. The results for 12
different energy bins are plotted in Fig.~\ref{zesWyk5}. One can see that in
each $W_{i}$ bin $\lambda_{\text{min}}(x_{\mathrm{cut}})$ can be approximated
by a constant, although for lower energies a slight increase for
$x_{\text{cut}}\geq0.1$ is present. Large error bars for higher energies are
due to the fact that $\chi_{W_{i}}^{2}$ gets flatter once $W_{i}$ is close to
$W_{\text{ref}}$.  To conclude that GS is well satisfied we have to
check whether the corresponding $\lambda_{\text{min}}(W_{i})$ for given $x_{\text{cut}}$ are energy
independent. This condition is satisfied if we restrict the energy range to
$W_{i}\geq33$ GeV.
Therefore the final number of data points used in the analysis in this Section is further reduced to 303.
In Fig.~\ref{Wchixcut}.a we plot the corresponding values of $\chi_{W_{i}}^{2}$.
One can see that up to $x_{\text{cut}}\simeq0.1$ the $\chi_{W_{i}}^{2}$'s are
smaller than $1.2$, and the rapid growth is seen for $x_{\text{cut}}>0.1$.

In order to get rid of the energy dependence we define yet another
chi-square function%
\begin{equation}
\tilde{\chi}^{2}(x_{\mathrm{cut}};\lambda)=\frac{1}{N_{x_{\text{cut}}}-1}%
\sum\limits_{i}\frac{(\lambda_{\mathrm{min}}\left(  W_{i},x_{\mathrm{cut}%
}\right)  -\lambda)^{2}}{\Delta\lambda_{\mathrm{min}}\left(  W_{i}%
,x_{\mathrm{cut}}\right)  ^{2}} \label{Wchitilde}%
\end{equation}
where the sum goes over eight energies $W_{i}=33,\,43,\ldots,159,\,268$ GeV. By
minimizing (\ref{Wchitilde}) with respect to $\lambda$ we obtain the best
value of $\lambda_{\mathrm{min}}\left(  W_{i},x_{\mathrm{cut}}\right)  $
averaged over all energies denoted by $\lambda_{\text{ave}%
}(x_{\text{cut}})$ which is plotted in Fig.~\ref{Wchixcut}.b. 
The error of $\lambda_{\text{ave}}$ is calculated again by
demanding that $(N_{x_{\text{cut}}}-1)\tilde{\chi}^{2}(x_{\mathrm{cut}%
};\lambda)$ changes by $1$ when $\lambda$ is varied around $\lambda
_{\text{ave}}$.  We see that $\lambda_{\text{ave}}(x_{\text{cut}})$ is
rather flat; large errors for small $x_{\text{cut}}$ are due to the small
number of points with small $x$'s. Looking at $\lambda_{\text{ave}%
}(x_{\text{cut}})$ one does not see any dramatic change for $x_{\text{cut}%
}\rightarrow1 $.  The values of $\tilde{\chi}^{2}(\lambda_{\text{ave}})$ fluctuate
around 1 with the highest value being smaller $2$ for all $x_{\rm cut}$ which 
is a clear sign of energy independence of  $\lambda_{\mathrm{min}}$.

Summarizing, we conclude that for energy binning GS  works
well up to $x_{\rm cut}=0.1$ where individual $\chi^2_{W_i}$'s start growing,
yelding
\begin{equation}
\lambda_{\text{en}}=\lambda_{\text{ave}}(x_{\text{cut}}=0.1)=0.343\pm 0.004.
\label{lenfinal}%
\end{equation}
where subscript "en" stands for energy binning. The error here is purely statistical.
Note, however,  that $\lambda_{\rm en}$ is larger than $\lambda_{\rm Bj}$ and
the difference is larger than the statistical errors. This difference may be used
to asses accuracy of our method. Had we used all energies, not excluding
$W_i$'s up to 33 GeV, we would get $\lambda_{\text{en}}=0.329\pm0.003$,
exactly as in the case of Bjorken binning.

\section{Summary and outlook}

\label{sumout}

In the present paper we have performed quantitative, model independent
analysis of the accuracy and the applicability domain of geometrical scaling
in deep inelastic $e^{+}p$ scattering. To this end we have chosen the most
recent compilation of the HERA data based on common analysis of ZEUS and H1
experiments \cite{HERAcombined}. We have tested the standard form of
geometrical scaling \cite{Stasto:2000er,GolecBiernat:1998js} given by the form
of the scaling variable $\tau$ (\ref{taudef}), assuming the constant value of
the exponent $\lambda$. In order to quantify the quality of GS we have
proposed two different approaches. In the first approach we have computed
ratios of cross-sections $F_{2}/Q^{2}$ for two different Bjorken $x$'s,
denoted as $x$ and $x_{\text{ref}}$, as functions of the scaling variable
$\tau$; such ratios should be equal to unity if GS is present. This allowed us
to extract the best value of the exponent $\lambda$, which we called
$\lambda_{\text{min}}(x,x_{\text{ref}}).$ In the region where GS is present
$\lambda_{\text{min}}(x,x_{\text{ref}})$ should not depend on $x$ and
$x_{\text{ref}}$, and moreover the corresponding $\chi^{2}$ should not
be too large.
In the second approach we have repeated essentially the same calculation but
for $\gamma^{\ast}p$ cross-sections treated as functions of the scattering
energy $W$ rather than $x$. This procedure is analogous to the one applied to
the $p_{\text{T}}$ spectra at the LHC. It requires, however, "rebinning" of the
data, which reduces the statistics and introduces uncontrollable errors, for
example the uncertainty  of $x$, and also in principle some uncontrollable
dependence on the choice of $W$ bins. Nevertheless, since $\gamma^{\ast}p$
cross-sections for all available energies overlap in $Q^{2}$ (or $\tau$)
nearly over whole kinematical range (which is not the case for the $x$
binning, see Figs. \ref{HERAxref} and \ref{HERAWref}), the analysis is much
more straightforward than in the case of the Bjorken $x$ binning.

\begin{figure}[t]
\centering
\includegraphics[width=7.5cm]{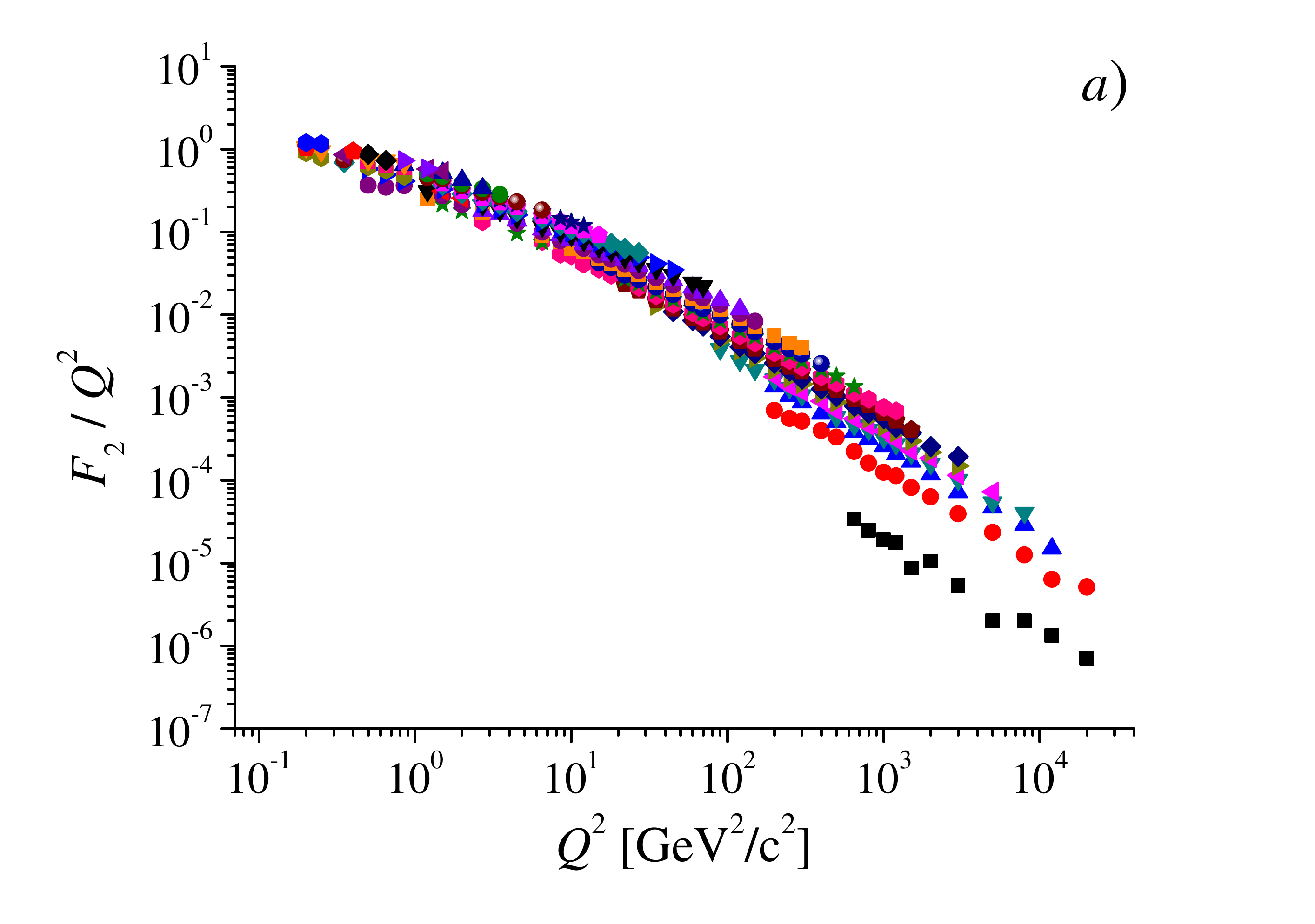}
\includegraphics[width=7.5cm]{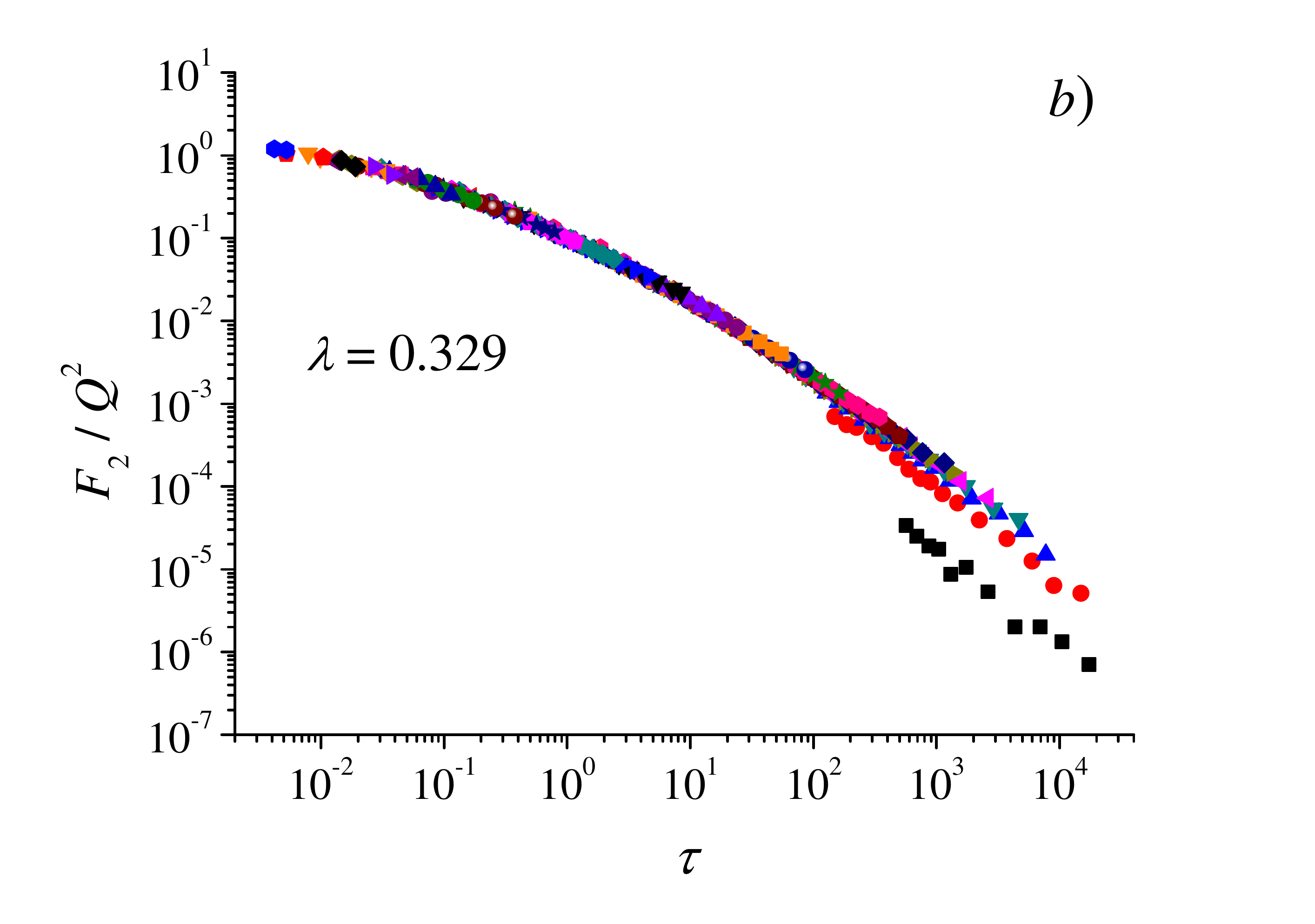}\caption{Geometrical scaling for the
Bjorken binning. Left: $\gamma^{\ast}p$ cross-sections $F_{2}/Q^{2}$ as
functions of $Q^{2}$ for fixed $x$. Different points correspond to different
Bjorken $x$'s. Right: the same but in function of scaling variable $\tau$ for
$\lambda=0.329$. Points in the right end of the plot correspond to large $x$'s
(due to kinematical correlation of the HERA phase space), and therefore show
explicitly violation of geometrical scaling.}%
\label{figGSinx}%
\end{figure}

\begin{figure}[t]
\centering
\includegraphics[width=7.5cm]{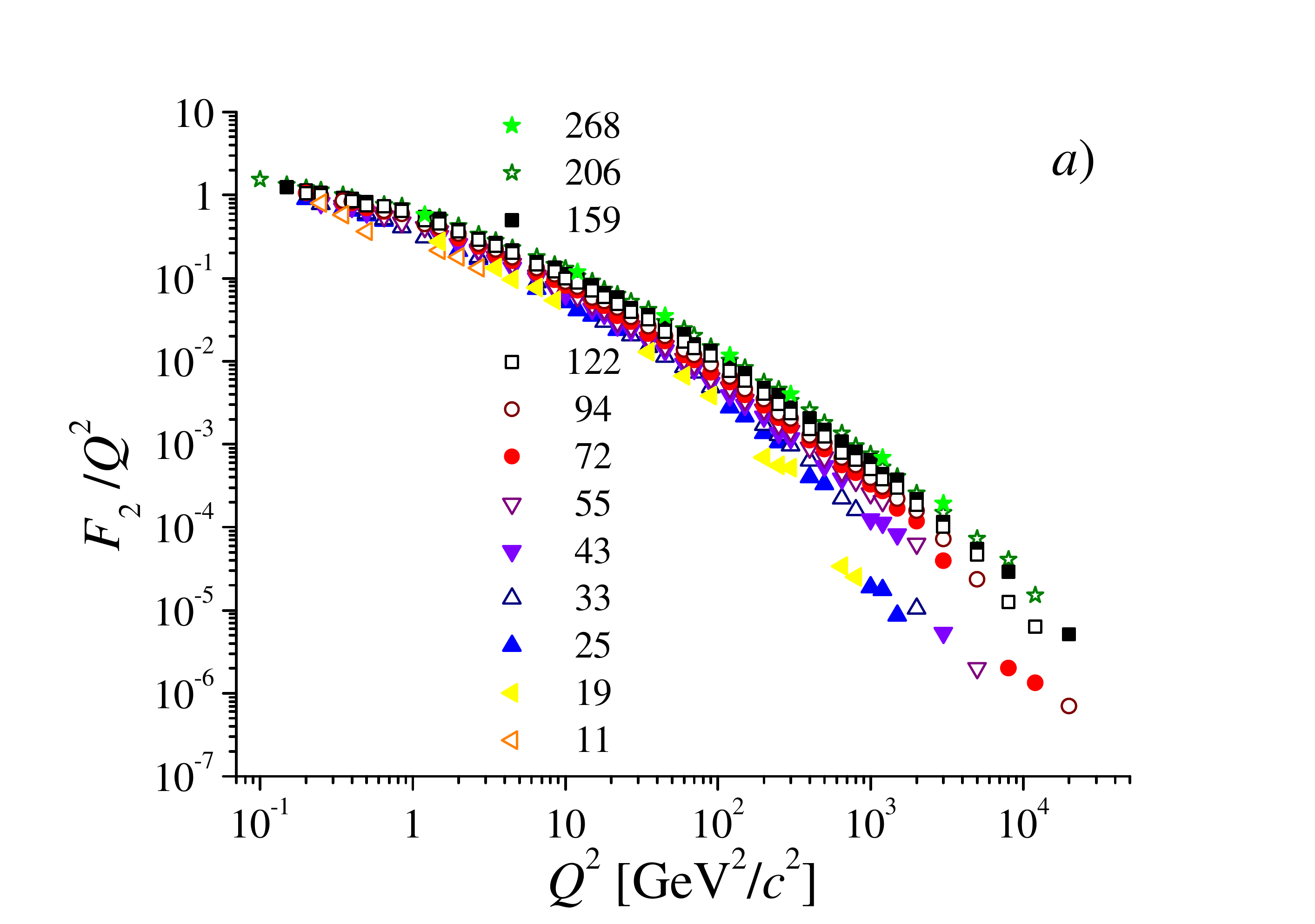}
\includegraphics[width=7.5cm]{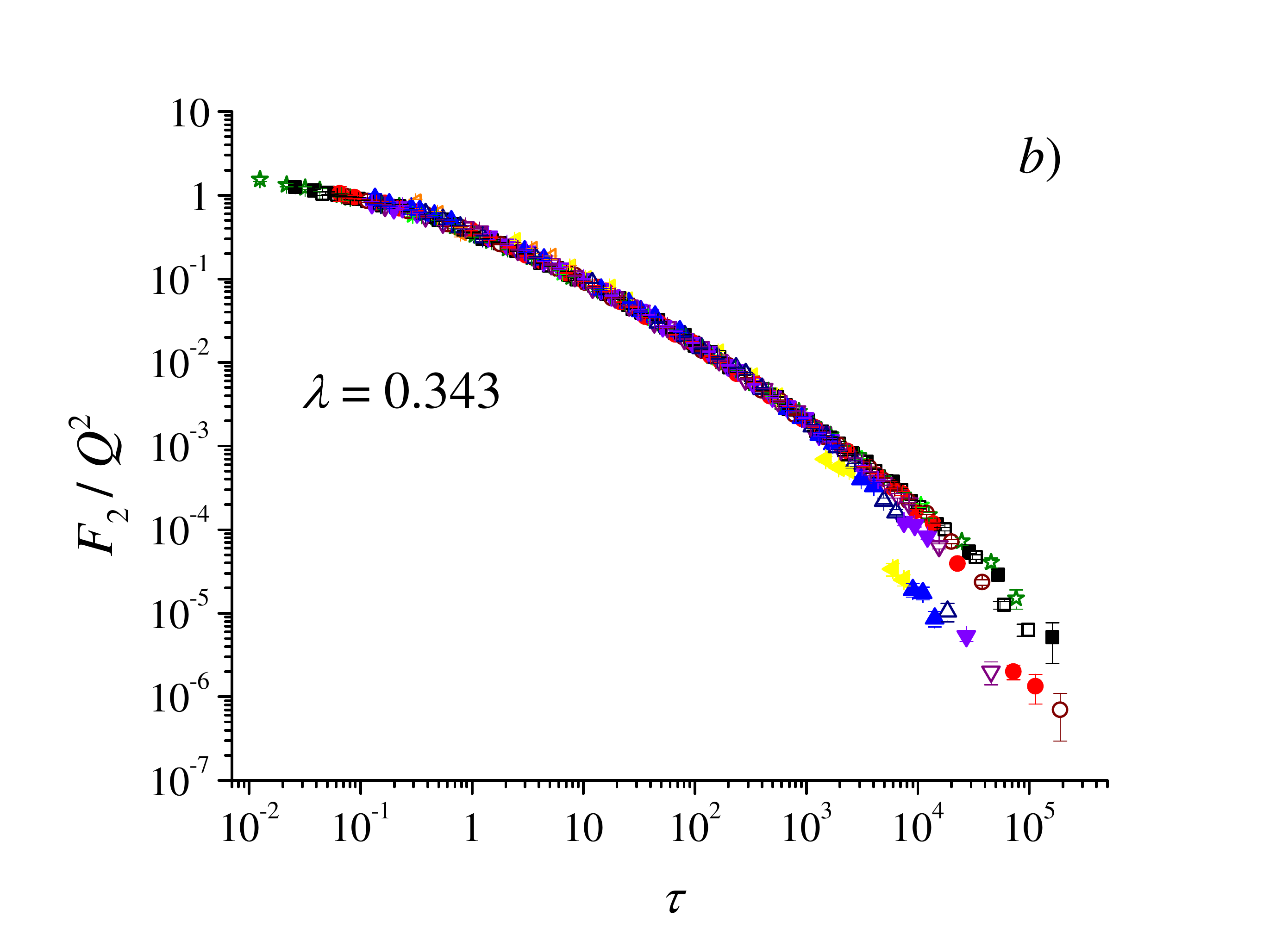}\caption{Geometrical scaling for the
energy binning. Left: $\gamma^{\ast}p$ cross-sections $F_{2}/Q^{2}$ as
functions of $Q^{2}$ for fixed $W$'s. Right: the same but in function of
scaling variable $\tau$ for $\lambda=0.343$. Points in the right end of the
plot correspond to large $x$'s and mostly small energies (due to kinematical
correlation of the HERA phase space), and therefore show explicitly violation
of geometrical scaling.}%
\label{figGSinW}%
\end{figure}

As mentioned in the Introduction we have assumed no error on $Q^2$, {\em i.e.}
$\Delta Q^{2}=0$. In Ref.~\cite{Stebel:2012ky} we have estimated 
$\Delta Q^{2}$ taking approximately half of the of the width of the $Q^{2}$ bin 
what should be considered as an upper bound of the real error. As a result the values
of different $\chi^2$ functions discussed in that paper have been underestimated.
However, since we never refer  to an absolute value of any given $\chi^2$, but
rather to the qualitative change of their behavior with  $x_{\rm cut}$ or $x_{\rm ref}$,
the conclusions concerning the validity of geometrical scaling do not depend 
much on actual values of $\Delta Q^2$ (with an obvious difference that theoretical error
$\delta$ can be safely neglected once errors of $Q^2$ are large). Qualitative behavior
of the $\chi^2$ functions is the same as in the present paper, although their values
are smaller and their behavior with $x_{\rm cut}$ or $x_{\rm ref}$ more smooth. 
The values of $\lambda$, however, are essentially unchanged.

Despite the differences between the Bjorken $x$ binning and the energy binning
the results in both cases are consistent and to some extent surprising. By
averaging $\lambda_{\text{min}}(x,x_{\text{ref}})$ or $\lambda_{\text{min}%
}(W,W_{\text{ref}})$ over the allowed regions of $x$ or $W$ respectively, and
by inspecting the corresponding $\chi^{2}$ values, as described in detail in
Sects. \ref{Bjxbinning} and \ref{Wbinning}, we have established that
geometrical scaling holds up to relatively large $x$'s of the order of $0.1$. 
The fact that
GS works up to large Bjorken $x$'s has been also observed in data analysis of
Ref.~\cite{Caola:2010cy}. 
The resulting exponents $\lambda$ given by Eqs. (\ref{lBjfinal}) and (\ref{lenfinal}) are slightly 
different since in the latter case we removed points with $W \leq 25$ GeV. 
Difference between those values shows systematic uncertainties arising from the change 
of binning.  In any case exponents $\lambda$ obtained in this paper are consistent with previous 
estimates and model calculations, and read
\begin{equation}
\lambda=0.32\div0.34.
\end{equation}

These results are suggestively illustrated in Figs.~\ref{figGSinx} and
\ref{figGSinW} where we plot $\gamma^{\ast}p$ cross-sections for both types of
binning used in this paper as functions of $Q^{2}$ and the scaling variable
$\tau$ (with $Q_{0}=1$ GeV/$c$ and $x_{0}=10^{-3}$ in Eq.(\ref{Qsat})). One
can see that in both cases the cross-sections fall on one universal curve -- a
clear sign of geometrical scaling. Some deviations of GS can be seen in the
lower right end of the plots. These points, however, correspond to large $x$'s
(due to the kinematical correlation of the HERA phase space, see Figs.
\ref{HERAxref}, \ref{HERAWref}), and therefore should not exhibit geometrical scaling.

The analysis performed here is essentially identical to the one presented in
in Ref.~\cite{Stebel:2012ky} with two important differences. Firstly, 
in Ref.~\cite{Stebel:2012ky} we have
used large errors of $Q^2$. Secondly, here, when
interpolating $\gamma^{\ast}p$ cross-section in $Q^{2}$ we use linear
interpolation in $\log Q^{2}$, rather than in $Q^{2}$ used in Ref.~\cite{Stebel:2012ky}. 
The latter
interpolation introduced large errors which were neglected in 
Ref.~\cite{Stebel:2012ky}. This resulted in somewhat different values of the
exponent $\lambda$. Qualitative picture, however, remains unchanged.

One of the questions addressed in \cite{Stebel:2012ky} was possible dependence
of $\lambda$ on $Q^{2}$. Applying the same method of dividing $\gamma^{\ast}p$
cross-sections of different Bjorken $x$'s (or scattering energies $W$) we have
been looking for the best exponent $\lambda$ which, in this case however,
depended on $Q^{2}$. Instead of having one parameter $\lambda$ we had
therefore a number of parameters $\lambda_{k}$ where $k$ runs over all $Q^{2}$
bins. By minimizing $\chi^{2}$ analogous to (\ref{chix1}) or (\ref{defchi2})
we did not observe any visible dependence of $\lambda$ on $Q^{2}$. This is
most probably due to the fact that the corresponding $\chi^{2}$ functions are
very flat and one can easily fall into some accidental minimum. We therefore
conclude that with the present experimental accuracy we have been unable to
find $Q^{2}$ dependence of $\lambda$ in a model independent way. It seems that
in this case one has to assume some Ansatz for $\lambda(Q^{2})$ depending on a
few variational parameters whose values can be found using the methods described in
this paper. Universality of these parameters would then constitute a criterion
for $Q^{2}$ dependence of $\lambda$. It is worthwhile to remark at this point that other
variants of the saturation scale have been discussed in the literature (for discussion  
see {\em e.g.} \cite{Gelis:2006bs} and references therein). Application of our
method to test the quality of GS for these  different scaling variables (including  
$Q^2$- dependent $\lambda$) is under preparation and will be presented elsewhere.

We have also analyzed separately $\gamma^{\ast}p$ cross-section for $e^{-}p$
scattering. The $e^{-}p$ data cover narrower range of $Q^{2}$ values and
therefore the smallest $x_{\mathrm{min}}=0.002$ is rather large as far as GS
is concerned. Nevertheless the results for $e^{-}p$ and $e^{+}p$ scattering
restricted to the same kinematical region are very similar except for some
small systematic shift of the exponent $\lambda$ towards higher values in the
$e^{-}p$ case \cite{Stebel:2012ky}.

Natural question concerning other deep inelastic scattering experiments
arises. Here we have $\mu p$ scattering experiments: EMC \cite{Aubert:1985fx}
with the smallest Bjorken $x_{\text{min}}=0.0175$, BCMDS
\cite{Benvenuti:1989rh} ($x_{\text{min}}=0.07$), NMC \cite{Arneodo:1996qe}
($x_{\text{min}}=0.008$), E665 \cite{Adams:1996gu} ($x_{\text{min}}%
=8\times10^{-4}$) and SLAC $e^{-}p$ experiment \cite{Whitlow:1991uw}
($x_{\text{min}}=0.063$). Let us remind here that the lowest Bjorken $x$ for
the combined HERA analysis is by far the smallest: $x_{\text{min}}%
=5.52\times10^{-6}$. This alone singles out HERA data as the best experimental
sample to look for geometrical scaling. Since, basing on our experience with
$e^{-}p$ data, we expect some possible systematic differences due to the
projectile used in different experiments, we have analyzed -- following the
steps of Sect. \ref{Bjxbinning} -- each of these experiments separately. We
have observed, indeed, rather large systematic differences between various
experiments in the overlapping region of Bjorken $x$'s (\emph{i.e.} for large
$x$) as far as $\left\langle \lambda_{\text{min}}\right\rangle $ and
$\left\langle \left\langle \lambda_{\text{min}}\right\rangle \right\rangle $
are concerned. However, due to large errors of $\left\langle \lambda
_{\text{min}}\right\rangle $ and $\left\langle \left\langle \lambda
_{\text{min}}\right\rangle \right\rangle $ it is difficult to quantify the
amount of violation of GS in these experiments. As far as E665 data is
concerned, we could see qualitative agreement with the combined HERA data,
although E665 data suffer from very large errors for $\left\langle
\lambda_{\text{min}}\right\rangle $ and $\left\langle \left\langle
\lambda_{\text{min}}\right\rangle \right\rangle $. More detailed analysis and
comparison of DIS experiments
\cite{Aubert:1985fx,Benvenuti:1989rh,Arneodo:1996qe,Adams:1996gu,
Whitlow:1991uw} with the combined HERA data \cite{HERAcombined} will be
presented elsewhere.

It is worthwhile mentioning that our results for the best value of the
exponent $\lambda$ are in good agreement with the quality factor
analysis of Ref.~\cite{Gelis:2006bs} $\lambda=0.321\pm0.056$ and with later
analysis of the combined HERA data only \cite{Royon:2010tz} $\lambda=0.31.$  
In both cases, however, the analysis was constrained to $x<0.01$ with some
additional cuts on $Q^{2}$ and $y$ which have not been applied in our study.

Obviously our method can be applied to test different forms of scaling
variable $\tau$ which follow from the QCD non-linear evolution equations
\cite{Gelis:2006bs,Beuf:2008bb,Royon:2010tz}. Finally, the question of the
possible existence of geometrical scaling for the charm cross-section
$F_{2}^{c}/Q^{2}$ \cite{Beuf:2008bb,Avsar:2007ht} where the charm quark mass
plays an important role may also be studied with the method proposed in this paper.

\section*{Acknowledgements}

MP would like to thank Robi Peschanski for discussion and for drawing his
attention to the quality factor studies of geometrical scaling and Gosta
Gustafson for pointing out the problem of the charm $F_{2}^{c}$ studied in
Ref.~\cite{Avsar:2007ht}. This work was supported by the Polish NCN grant 2011/01/B/ST2/00492.

\end{document}